\documentclass[12pt]{iopart}
\usepackage{iopams}
\expandafter\let\csname equation*\endcsname\relax 
\expandafter\let\csname endequation*\endcsname\relax

\usepackage{amsmath}
\usepackage{alphalph, relsize}

\usepackage[T1]{fontenc} 
\usepackage{ulem}
\usepackage{ragged2e} 

\usepackage[outline]{contour}
\usepackage{xcolor,graphicx}
\usepackage{caption}
\usepackage{subcaption}
\usepackage{bm}

\usepackage{tensor, enumerate, bbold }
\usepackage{array}
\usepackage[perpage]{footmisc}

\usepackage{color,marvosym}
\usepackage{graphicx} 
\usepackage{comment}
\usepackage{cancel}

\definecolor{darkred}{rgb}{0.5,0,0}
\definecolor{darkgreen}{rgb}{0,0.5,0}
\definecolor{darkblue}{rgb}{0,0,0.5}

\usepackage{hyperref}
\hypersetup{colorlinks,
linkcolor=darkblue,
filecolor=darkgreen,
urlcolor=darkblue,
citecolor=darkblue }
\newcommand{\const}{\textrm{const}}
\newcommand{\UD}[2]{\ensuremath{^{#1}_{\phantom{#1} #2}}}

\newcommand{\UDD}[3]{\ensuremath{^{#1}_{\phantom{#1} #2 #3}}}
\newcommand{\UDDD}[4]{\ensuremath{^{#1}_{\phantom{#1} #2 #3 #4}}}

\newcommand{\beq}{\begin{equation}}
\newcommand{\eeq}{\end{equation}}

\newcommand{\bwt}{\begin{widetext}}
\newcommand{\ewt}{\end{widetext}}

\newcommand{\bea}{\begin{eqnarray}}
\newcommand{\eea}{\end{eqnarray}}
\newcommand{\bean}{\begin{eqnarray*}}
\newcommand{\eean}{\end{eqnarray*}}
\newcommand{\bit}{\begin{itemize}}
\newcommand{\eit}{\end{itemize}}
\newcommand{\bfi}{\begin{figure}}
\newcommand{\efi}{\end{figure}}
\newcommand{\bfic}{\begin{figure*}}
\newcommand{\efic}{\end{figure*}}
\newcommand{\bce}{\begin{center}}
\newcommand{\ece}{\end{center}}
\newcommand{\bt}{\begin{table}}
\newcommand{\et}{\end{table}}
\newcommand{\btb}{\begin{tabular}}
\newcommand{\etb}{\end{tabular}}

\newcommand{\calE}{\ensuremath{\mathcal{E}}}
\newcommand{\calO}{\ensuremath{\mathcal{O}}}
\newcommand{\calS}{\ensuremath{\mathcal{S}}}

\newcommand{\calW}{\ensuremath{\mathcal{W}}}

\newcommand{\WXL}{\ensuremath{\mathcal{W}_{\rm XL}}}
\newcommand{\WXX}{\ensuremath{\mathcal{W}_{\rm XX}}}
\newcommand{\WLL}{\ensuremath{\mathcal{W}_{\rm LL}}}
\newcommand{\WLX}{\ensuremath{\mathcal{W}_{\rm LX}}}

\newcommand{\stuff}{\mathcal{H}_0^2\Omega_{\rm m_0}}

\newcommand{\qed}{\nobreak \ifvmode \relax \else
      \ifdim\lastskip<1.5em \hskip-\lastskip
      \hskip1.5em plus0em minus0.5em \fi \nobreak
      \vrule height0.75em width0.5em depth0.25em\fi}
 
\begin{document}

\title{BiGONLight: light propagation with bi-local operators in Numerical Relativity}
\author{Michele Grasso and Eleonora Villa}
\address{Center for Theoretical Physics Polish Academy of Sciences
Al. Lotnik\'ow 32/46, 02-668 Warsaw Poland.}
\ead{grasso@cft.edu.pl, villa@cft.edu.pl}
\vspace{10pt}
\begin{indented}
\item[]\date{\today}
\end{indented}

\begin{abstract}
{\tt BiGONLight}, \textbf{Bi}local \textbf{G}eodesic \textbf{O}perators framework for \textbf{N}umerical \textbf{Light} propagation, is a new tool for light propagation in Numerical Relativity. The package implements the Bi-local Geodesic Operators formalism, a new framework for light propagation in General Relativity. With {\tt BiGONLight} it is possible to extract observables such as angular diameter distance, luminosity distance, magnification as well as new real-time observables like parallax and redshift drift within the same computation. As test-bed for our code we consider two cosmological models, the $\Lambda$CDM and the inhomogeneous Szekeres model, and a simulated dust universe. All our tests show an excellent agreement with known results.

\end{abstract}

\submitto{\CQG}
\maketitle

\section{Introduction}
Electromagnetic and gravitational radiation are the primary means by which cosmologists and astronomers try to learn about the structure and the evolution of the Universe.  All the information we receive from distant objects is inferred from these signals which are affected by the presence of cosmic structures between the source and the observer. These effects are accumulated along the line of sight and they modify the perception of the observer which e.g. can receive the image of the source as if it were in a different position on the sky or with a distorted shape or measure a redshift in the source light spectrum. 
Actually, most of the information comes from these distortions and they will be measured with unprecedented precision in a wider range of scales and redshift by the next generation of galaxy survey and Cosmic Microwave Background experiments\footnote{{\color{blue}{\tt {https://www.skatelescope.org}}}, {\color{blue}{\tt {https://www.euclid-ec.org}}}, {\color{blue}{\tt {https://www.lsst.org}}}, {\color{blue}{\tt {http://litebird.jp/eng/}}}, {\color{blue}{\tt {https://www.jpl.nasa.gov/missions/spherex}}}}. On top of that, thanks to the improvements in the modern experimental apparatus, nowadays we will be able of measuring small temporal changes of those effects, the so-called optical drift effects. They may provide important and new information about the Universe structure and evolution, marking the beginning of Real-time Cosmology, \cite{Quercellini:2010zr}. In order to make the most from this revolution in observational astronomy and cosmology, the same accuracy is required in the theoretical predictions and interpretations of what we measure. This tough task requires much effort from two sides. At one hand the most recent progress on cosmological dynamics are represented by  general-relativistic simulations of cosmic structures with no assumed symmetries employing exact solutions of Einstein equations, \cite{loffler2012einstein,Giblin:2015vwq,Bentivegna:2016stg,macpherson2017,Clesse:2017jjp,East:2017qmk,Daverio:2019gql}, or approximated treatments, \cite{adamek2016general,barrera2020gramses}. The common ambitious aim is to have a description valid from large to small scales which accounts for relativistic effects. 
On the other hand relativistic effects in the non-linear regime have recently become to be investigated with relativistic simulations in galaxy clustering and lensing observables, and the Hubble diagram, see e.g. \cite{Borzyszkowski:2017ayl, Zhu:2017jfl, Giblin:2017ezj, Breton:2018wzk, Adamek:2018rru, Beutler:2020evf, Lepori:2020ifz, Guandalin:2020snp, Lepori:2021lck} and refs. therein. 

From the point of view of the basic theory of light propagation a new approach was presented in \cite{grasso2019BGO}. The key ingredients of this new formulation are the Bi-local Geodesic Operators (BGO) which constitute the map from the portion of spacetime occupied by the observer to that occupied by the source and give the full description of the distortion of light rays in between. The main advantage of the BGO formalism is that it provides a unified framework to compute all optical observables, namely those inferred from gravitational lensing effects, like e.g. magnification, shear, and angular diameter distance. The novelty is that source(s) and observer(s) are allowed to move and therefore the observables originated by the variation in time and the motion of sources and observers, like e.g. parallax, and redshift and position drifts, are automatically included, see \cite{grasso2019BGO}. In addition, by having the observables written in terms of the BGO, it is easy to disentangle the contributions of the spacetime curvature from those due to observer and source motion and construct specific new probes, e.g. for the curvature as the distance slip investigated in \cite{ korzynski2020geometric} in the $\Lambda$CDM cosmology.

In this paper we present {\tt BiGONLight}\footnote{The package is publicly available at {\color{blue}{\tt {https://github.com/MicGrasso/bigonlight.git}}}.}, \textbf{Bi}local \textbf{G}eodesic \textbf{O}perators framework for \textbf{N}umerical \textbf{Light} propagation, a \texttt{Mathematica} package developed for extracting observables from numerically generated spacetimes using the BGO formalism. The principal aim of the package is to provide a unified procedure to calculate multiple observables in Numerical Relativity: this is guaranteed since, once that the BGO are determined from the output metric of a numerical simulation, a all set of observables are obtained within the same computation. In order to be compatible with the majority of the codes in Numerical Cosmology, {\tt BiGONLight} encodes the BGO formalism in $3+1$ form and it uses the \texttt{Mathematica} powerful symbolic algebra manipulation and precision control options. 

The paper is organized as follows: in Sec.~\ref{sec:light}, we give the fundamentals of light propagation and its formulation in terms of Bi-local Geodesic Operators. In Sec.~\ref{sec:3+1} we  present {\tt BiGONLight} and the equations to compute the BGO in $3+1$ form encoded into the package. The recipe to compute observables in numerical relativity using {\tt BiGONLight} and the expressions of the observables in terms of BGO are given in Sec.~\ref{sec:obs}. The last section of this paper, Sec.~\ref{sec:test}, is dedicated to the code tests, which are performed in the following three cosmological models: $\Lambda$CDM (see Sec.~\ref{sec:LCDM}), Szekeres (see Sec.~\ref{sec:Szekeres}), and a numerically evolved dust universe (see Sec.~\ref{sec:ET}). We draw our conclusions in Sec.~\ref{sec:concl}.\\

Notation: Greek indices ($\alpha, \beta, ...$) run from 0 to 3, while Latin indices ($i,j, ...$) run from 1 to 3 and refer to spatial coordinates only. Latin indices ($A,B, ...$) run from 1 to 2. Tensors and bitensors expressed in a semi-null frame are denoted using boldface indices: Greek boldface indices ($\boldsymbol{\alpha}, \boldsymbol{\beta}, ...$) run from 0 to 3, Latin boldface indices ($\mathbf{a}, \mathbf{b}, ...$) run from 1 to 3 and capital Latin boldface indices ($\mathbf{A}, \mathbf{B}, ...$) run from 1 to 2. Latin tilded indices ($\tilde{a}, \tilde{b}, ...$), running from $0$ to $7$, denote indices for the components of the $8 \times 8$ BGO matrix $\mathcal{W}$. A dot denotes derivative with respect to conformal time. Quantities with a subscript $0$ are meant to be evaluated at present, whereas the subscript $``{\rm in}"$ indicates the initial time. Quantities with a subscript $\cal S$ (or $\cal O$) are meant to be evaluated at the source (observer) position.


\section{Light propagation and the bi-local geodesic operators}
\label{sec:light}
\begin{figure}[h]
\centering
\includegraphics[width=0.8\textwidth]{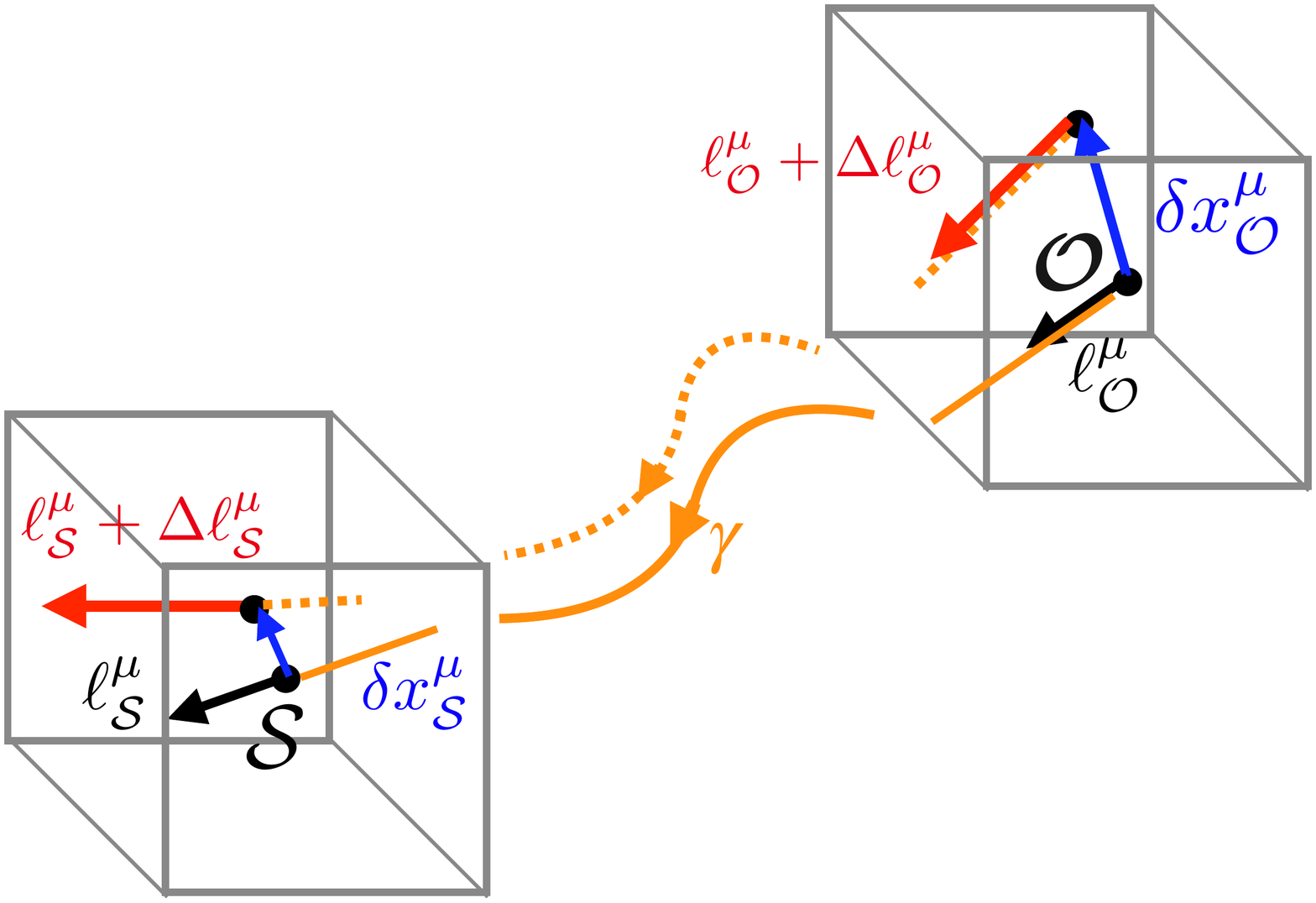}
       \caption{When the observations are repeated over time, the observer and the source may change their positions. This implies that the null geodesic connecting $\mathcal{O}$ to $\mathcal{S}$ has to change its path accordingly. This situation is described with the deviations of the positions $\delta x^{\mu}$ and of the tangent vectors $\Delta \ell^{\mu}$ at the two new extremes $\mathcal{S}$ and $\mathcal{O}$.}
        \label{fig:geom_setup}
\end{figure}

Let us start by considering an observer $\mathcal{O}$ and a source $\mathcal{S}$ separated by a large distance and moving freely along their time-like worldlines. Naming $\mathcal{N}_{\mathcal{O}}$ and $\mathcal{N}_{\mathcal{S}}$ the regions of the spacetime in which the observer and the source are moving, we assume that $\mathcal{N}_{\mathcal{O}}$ and $\mathcal{N}_{\mathcal{S}}$ are causally connected, so that any signal emitted by $\mathcal{S}$ is received by $\mathcal{O}$ at any later time. We also assume that the typical length scales of $\mathcal{N}_{\mathcal{O}}$ and $\mathcal{N}_{\mathcal{S}}$ are small compared to the distance between them.
Within the geometric optics approximation, we can describe the signal as moving along null geodesics $\gamma(\lambda)$ connecting the source and the observer, such that $\gamma(\lambda_{\mathcal{S}})=x^{\mu}_{\mathcal{S}}$ and $\gamma(\lambda_{\mathcal{O}})=x^{\mu}_{\mathcal{O}}$, where $x^{\mu}_{\mathcal{O}}$ and $x^{\mu}_{\mathcal{S}}$ are the observer's and source's positions, respectively. 
The curve $\gamma$ is given by the geodesic equation
\begin{equation}
\ell^{\sigma} \nabla_{\sigma}\ell^{\mu}=\frac{D}{D \lambda}\ell^{\mu}=0\, ,
\label{eq:geod_eq}
\end{equation}
where $\ell^{\mu}$ is the tangent vector, $\lambda$ is the affine parameter spanning the geodesic $\gamma$, $\frac{D}{D \lambda}\equiv\ell^{\sigma}\nabla_{\sigma}$ is the covariant derivative along $\gamma$. A solution of Eq.~\eqref{eq:geod_eq} is specified by the initial position and the initial tangent vector. In astronomy, we are the only observer performing every measurement, thus it is natural to set the initial conditions ($x^{\mu}_{\mathcal{O}}$, $\ell^{\mu}_{\mathcal{O}}$) at the observer location and to trace the geodesic back to the source. Now, if the observer is displaced by $\delta x^{\mu}_{\mathcal{O}}$, a new geodesic connects $\mathcal{O}$ and $\mathcal{S}$, and it is characterized by the new initial conditions ($x^{\mu}_{\mathcal{O}}+\delta x^{\mu}_{\mathcal{O}}$, $\ell^{\mu}_{\mathcal{O}} + \Delta \ell^{\mu}_{\mathcal{O}}$), see Fig.\ref{fig:geom_setup}, where we define $\Delta \ell^{\mu}_{\mathcal{O}}$ as the covariant deviation of the tangent vector $\ell^{\mu}_{\mathcal{O}}$, namely
\begin{equation}
\Delta \ell^{\mu}_{\mathcal{O}}=\delta \ell^{\mu}_{\mathcal{O}}+\Gamma\UDD{\mu}{\alpha}{\beta}(x_\mathcal{O})\ell^{\alpha}_{\mathcal{O}} \delta x^{\beta}_{\mathcal{O}}\, .
\end{equation} 

The deviations ($\delta x^{\mu}_{\mathcal{O}}$, $\Delta \ell^{\mu}_{\mathcal{O}}$) can be used to parametrize a family of null geodesics and they propagate according to the Geodesic Deviation Equation (GDE)
\begin{equation}
\left\{ \begin{matrix}
\dfrac{D}{D \lambda} \delta x^{\mu}= \Delta \ell^{\mu}\\
\\
\dfrac{D}{D \lambda} \Delta \ell^{\mu}= R\UDDD{\mu}{\ell}{\ell}{\nu} \delta x^{\nu}
\end{matrix}\right.\, 
\label{eq:GDE_deviations}
\end{equation}
with initial conditions
\begin{align}
\delta x^{\mu}(\lambda_{\mathcal{O}})&=\delta x^{\mu}_{\mathcal{O}}\\
\Delta \ell^{\mu}(\lambda_{\mathcal{O}})&= \Delta \ell^{\mu}_{\mathcal{O}}\, ,
\end{align}
provided that all the geodesics of the family stay close enough to $\gamma$\footnote{Here the GDE is presented as a system of two first-order ODE. The reader may find more familiar the following form of the GDE as a second order ODE $\frac{D^2}{D \lambda^2}\xi^{\mu}=R^{\mu}_{\ \  \alpha \beta \nu} \ell^{\alpha}\ell^{\beta}\xi^{\mu}$, where $\xi^{\nu}=\delta x^{\nu}$ and $\frac{D}{D \lambda}\xi^{\nu}=\Delta \ell^{\nu}$. In general, the GDE describes the deviation between any two infinity close  null-like, time-like or space-like geodesics.}.
In Eq. \eqref{eq:GDE_deviations} $ R\UDDD{\mu}{\ell}{\ell}{\nu}$ is the short-hand for the optical tidal matrix $ R\UDDD{\mu}{\ell}{\ell}{\nu}\equiv R\UDDD{\mu}{\alpha}{\beta}{\nu}\ell^{\alpha}\ell^{\beta}$. The standard procedure to solve the GDE \eqref{eq:GDE_deviations} was introduced by Sachs in 1961,  \cite{Sachs1961}. The key quantity of this approach is the deformation of the light bundle's cross section, i.e. the projection of the beam on a 2D screen spanned by two orthonormal vectors orthogonal to $\ell^{\mu}$ and parallelly transported along $\gamma$ (the Sachs basis). The GDE is projected onto this basis and then re-written in terms of the deformation matrix, a $2 \times 2$ matrix containing the expansion and the shear: the GDE recasted this way gives the well-known Sachs equations, see \cite{perlick2004lensing} for a complete introduction to the Sachs formalism. 
Despite its great success, the Sachs formalism describes only momentary observations, so that it can not be used to take into account what happens when the observations are repeated in time and/or the observer and the source move. Obviously, this can be overcome by considering a new fiducial geodesic at some later time and solving again the Sachs equations but this procedure can be very complicated in some cases. 
In this paper we will use a different approach, which accounts also for these  situations in a unified framework, based on the \textit{Bi-local Geodesic Operators} (BGO), introduced in \cite{grasso2019BGO} and summarised below.

Let us start by noticing that the GDE \eqref{eq:GDE_deviations} defines a linear, bijective map $\mathcal{W}(\calS, \calO)$ between its solutions ($\delta x^{\mu}_{\mathcal{S}}$, $\Delta \ell^{\mu}_{\mathcal{S}}$) and initial conditions ($\delta x^{\mu}_{\mathcal{O}}$, $\Delta \ell^{\mu}_{\mathcal{O}}$), namely
\begin{equation}
\begin{array}{l}
\delta x^{\mu}_{\mathcal{S}}= \WXX{}\UD{\mu}{\nu} \delta x^{\nu}_{\mathcal{O}} + \WXL{}\UD{\mu}{\nu} \Delta \ell^{\nu}_{\mathcal{O}}\\
\Delta \ell^{\mu}_{\mathcal{S}}= \WLX{}\UD{\mu}{\nu} \delta x^{\nu}_{\mathcal{O}} + \WLL{}\UD{\mu}{\nu} \Delta \ell^{\nu}_{\mathcal{O}}\, ,
\end{array}
\label{eq:deltax_deltal_BGO}
\end{equation}
assuming the optical tidal matrix $R\UDDD{\mu}{\ell}{\ell}{\nu}$ to be a smooth tensor field. 
The bi-tensors $\WXX$, $\WXL$, $\WLX$, $\WLL$ acting from $\mathcal{O}$ to $\mathcal{S}$ are called \textit{bi-local geodesic operators}.
Equation~\eqref{eq:deltax_deltal_BGO} can then be written in the more compact form
\begin{equation}
\left(\begin{array}{l}
\delta x_{\mathcal{S}}\\
\Delta \ell_{\mathcal{S}}
\end{array}\right)=\begin{pmatrix} 
\WXX & \WXL\\
\WLX & \WLL\\
\end{pmatrix}\left(\begin{array}{l}
\delta x_{\mathcal{O}}\\
\Delta \ell_{\mathcal{O}}
\end{array}\right)=\mathcal{W}(\mathcal{S}, \mathcal{O})\left(\begin{array}{l}
\delta x_{\mathcal{O}}\\
\Delta \ell_{\mathcal{O}}
\end{array}\right),
\label{eq:compact_W(S,O)}
\end{equation}
where $\mathcal{W}(\mathcal{S}, \mathcal{O})$ is the $8 \times 8$ Wro\'nski matrix of the GDE acting from $\mathcal{O}$ to $\mathcal{S}$.
By inserting~\eqref{eq:compact_W(S,O)} in the GDE \eqref{eq:GDE_deviations}, we obtain the propagation equation for the BGO
\begin{equation}
\dfrac{D}{D \lambda} \mathcal{W}=\begin{pmatrix}
0 & \mathbb{1}\\
R\UDDD{\mu}{\ell}{\ell}{\nu} & 0
\end{pmatrix} \mathcal{W}
\label{eq:GDE_for_BGO}
\end{equation}
with initial conditions
\begin{equation}
\mathcal{W} \left|_{\mathcal{O}} \right. = \begin{pmatrix}
\mathbb{1}_{4 \times 4} & 0\\
0 & \mathbb{1}_{4 \times 4}
\end{pmatrix}\, .
\end{equation}
The BGO $\mathcal{W}$ is a symplectic mapping, as firstly noticed in \cite{Uzun:2018yes}, since Eq.~\eqref{eq:GDE_for_BGO} for null geodesics can be formulated as a Hamiltonian system, and it satisfies the properties
\begin{align}
\mathcal{W}(\mathcal{O}, \mathcal{S})&=\mathcal{W}^{-1}(\mathcal{S}, \mathcal{O}) \label{eq:BGO_property1}\\
\mathcal{W}(\mathcal{S}, \mathcal{O})&=\mathcal{W}(\mathcal{S}, p_{\lambda}) \, \mathcal{W}(p_{\lambda}, \mathcal{O})\, ,
\label{eq:BGO_property2}
\end{align}
with $p_{\lambda}$ an arbitrary point on the fiducial geodesic $\gamma$.
The usual set up is specified by giving the initial conditions at $\mathcal{O}$ and by integrating the GDE backward in time, up to the source $\mathcal{S}$. The physical motivation is of course that every measurement is done from the observer position.
The BGO $\mathcal{W}(p_{\lambda},\mathcal{O})$ obtained by integrating Eq.~\eqref{eq:GDE_for_BGO} backwards connect the observer with an arbitrary point $p_{\lambda}$ on the geodesic, ending at the source  $p_{\lambda_{\mathcal{S}}}=\mathcal{S}$. 
Although it is natural to study light propagation this way, there are circumstances in which it is more convenient to think forward in time, namely to integrate Eq.~\eqref{eq:GDE_for_BGO} from the source to $p_{\lambda}$, ending at the observer $p_{\lambda_{\mathcal{O}}}=\mathcal{O}$, and obtain $\mathcal{W}(p_{\lambda},\mathcal{S})$. To this second case belong all the simulations of cosmological dynamics, in which the Einstein equations are solved forward in time, with initial conditions given at the end of inflation. It follows that forward integration of light propagation, on-the-fly with the simulation of the spacetime dynamics, would be a cost-efficient and time-saving methodology instead of using the natural approach for light propagation in post-processing. 
These two opposing procedures find a meeting point within the BGO framework, which provides a relatively easy way to transform from forward integrated $\mathcal{W}(p_{\lambda},\mathcal{S})$ to backward integrated $\mathcal{W}(p_{\lambda},\mathcal{O})$. The transformation follows from the BGO properties and it is found by multiplying Eq.~\eqref{eq:BGO_property2} by $\calW^{-1} (\calS,p_{\lambda})$ from the left and using Eq.~\eqref{eq:BGO_property1} to obtain\footnote{In the last equality of Eq.~\eqref{eq:W(p,o)_from_W(p,s)} we used Eq.~\eqref{eq:BGO_property1}. 
 The symplectic property is useful for getting the inverse matrix, see Sec.~\ref{sec:W3+1}.}
\begin{align}
\calW (p_{\lambda},\calO)&=\calW^{-1} (\calS,p_{\lambda})\,  \calW (\calS,\calO) \nonumber \\
&=\calW (p_{\lambda},\calS)\, \calW^{-1} (\calO, \calS) \, .
\label{eq:W(p,o)_from_W(p,s)}
\end{align}
We will give the explicit transformation rules for all the BGO components in Sec.~\ref{sec:W3+1}.

Let us finally recall that the most important feature of the BGO formalism is that it provides a unified framework to compute optical observables, including also real-time ones, like e.g. parallax, and redshift and position drifts, \cite{grasso2019BGO}. In Sec.~\ref{sec:obs} we will give the explicit expressions of the observables considered in this work in terms of the BGO.


\section{{\tt BiGONLight}: light propagation with the BGO in Numerical Relativity}
\label{sec:3+1}
The BGO framework finds a natural application in studying light propagation in Numerical Relativity, since it allows to obtain multiple optical observables within the same calculation and this is easily adaptable to perform light propagation on-the-fly with a simulation of relativistic dynamics.
For this purpose, we develop {\tt BiGONLight} (\textbf{Bi}-local \textbf{G}eodesic \textbf{O}perator framework for \textbf{N}umerical \textbf{Light} propagation), a {\tt Mathematica} package which simulates light propagation in Numerical Relativity within the BGO formalism, described in Sec.~\ref{sec:light}. It is publicly available at the repository {\color{blue}{\tt {https://github.com/MicGrasso/bigonlight.git}}} and it works as an external library that can be called inside a {\tt Mathamatica} notebook.  \texttt{Mathematica} provides a large variety of numerical methods that can be used and customized to adapt them to the particular problem.  This is exploited in the functions {\tt SolveGeodesic[]}, {\tt SolveEnergy[]}, {\tt PTransportedFrame[]} and {\tt SolveBGO[]} in which the user can choose the numerical methods used to solve the system of ODE\footnote{The user can choose between three different methods: $``RK''$ a $4^{th}$ order Runge-Kutta method, $``A''$ which lets {\tt Mathematica} to decide what is the best method to use, and $``SS''$ denoting the Stiffness-Switching method, which allows to switch between implicit and explicit methods to resolve stiff problems.}. Another useful feature is the \texttt{Mathematica}'s precision control options, which allows the user to set the  precision and accuracy of numerical result through the commands \texttt{WorkingPrecision}, \texttt{SetPrecision} and \texttt{SetAccuracy}. 
In the following we give a detailed description of the equations for light propagation and the procedure to obtain the BGO implemented in the code. We dedicate a separate section, Sec.~\ref{sec:obs}, to the computation of the observables once the BGO are known. 

The required input for {\tt BiGONLight} are the specetime metric and the source/observer kinematics, which can be provided in two different ways: (i) from the analytical expression of the metric $g_{\mu \nu}$ and the emitter/observer four-velocities ($u^{\mu}_{\mathcal{S}}$, $u^{\mu}_{\mathcal{O}}$) and four-accelerations ($w^{\mu}_{\mathcal{S}}$, $w^{\mu}_{\mathcal{O}}$) for some exact model, and (ii) from the output of a relativistic numerical simulation of the spacetime dynamics.
A large variety of numerical codes used in cosmology and astrophysics employs the 3+1 formalism to solve the Einstein equations and simulate full-GR dynamical systems. To be compatible with the numerical output generated by these codes, in {\tt BiGONLight} we recasted the BGO formalism in the 3+1 form.
In the following we summarise some basic definitions of the 3+1 formalism and report the 3+1 version of all the equations for light propagation used to compute the BGO. For comprehensive references of the 3+1 formalism see e.g. \cite{Smarr:1977uf, alcubierre2008introduction, baumgarte2010numerical, gourgoulhon20123+}.

The procedure of splitting a four-dimensional spacetime $(\mathcal{M},g_{\mu \nu})$ into its $3+1$ form, the so-called ADM formalism, was introduced by Arnowitt, Deser and Misner in \cite{arnowitt1959}. It is constructed by foliating a four-dimensional manifold $\mathcal{M}$ with a family of 3D space-like hypersurfaces $\Sigma_t$, labelled by a monotonic function $t$, such that $t=\const$ on each slice. This space-like foliation defines the time-like vector field $n^{\mu}$, which is orthonormal to the slices and it can be regarded as the four-velocity of Eulerian observes. 
The geometry on each hypersurface is described by the following quantities:
\begin{itemize}
\item the induced metric $\gamma_{\mu \nu}$ is defined as the covariant form of the orthogonal projector onto the slices 
\begin{equation}
\gamma_{\mu \nu}=g_{\mu \nu} + n_{\mu}n_{\nu}
\end{equation}
and it is used to measure proper distances on $\Sigma_t$;
\item the covariant derivative on the slice
\begin{equation}
D_{\nu}V^{\mu}=\gamma\UD{\sigma}{\nu}\gamma\UD{\mu}{\rho}\nabla_{\sigma}V^{\rho} \, ,
\label{eq:3D_covD}
\end{equation}
which is written in terms of the 3D Christoffel symbol ${}^{(3)}\Gamma^{k}_{\ i j}=\frac{1}{2}\gamma^{k l}(\frac{\partial \gamma_{l j}}{\partial x^i}+\frac{\partial \gamma_{i l}}{\partial x^j}-\frac{\partial \gamma_{i j}}{\partial x^l})$, once a coordinate system ${x^i}$ on $\Sigma_{t}$ is introduced;
\item the extrinsic curvature $K_{\mu \nu}$ defined as
\begin{equation}
K_{\mu \nu}=-\gamma\UD{\sigma}{\mu}\gamma\UD{\rho}{\nu}\nabla_{\sigma}n_{\rho}\, ,
\label{eq:K_def}
\end{equation}
which represents the curvature of the 3D hypersurfaces with respect to the 4D embedding spacetime.
\end{itemize}
A natural choice for a coordinate system $x^{\mu}$ is the one adapted to the foliation: the corresponding reference frame is such that the three vectors $\partial^{\mu}_i=\left(\frac{\partial}{\partial x^i}\right)^{\mu}$ are tangent to the hypersurface while $\partial^{\mu}_0=\left(\frac{\partial}{\partial t}\right)^{\mu}$ is transverse to it. In particular, the time-like vector field $\left(\frac{\partial}{\partial t}\right)^{\mu}$ is tangent to a congruence of world-lines of coordinate observers and it is given by
\begin{equation}
\left(\dfrac{\partial}{\partial t}\right)^{\mu}=\alpha n^{\mu}+\beta^{\mu}\, ,
\end{equation}
where $\alpha$ is the lapse function, which measures the proper time of the Eulerian observers, and $\beta^{\mu}$ is the  shift vector, which quantifies the displacement on $\Sigma_t$ of the coordinate observer $\left(\frac{\partial}{\partial t}\right)^{\mu}$ with respect to the Eulerian observer $n^{\mu}$. 
The components of the normal vector and the metric $g_{\mu \nu}$ in the adapted coordinates are written with the lapse, the shift and the induced metric as
\begin{equation}
n^{\mu}=(\dfrac{1}{\alpha},-\dfrac{\beta^i}{\alpha})\, ,
\label{eq:nu_nd}
\end{equation}
and
\begin{equation}
g_{\mu \nu}=\begin{pmatrix}
\beta_{i}\beta^i-\alpha^2 &~ \beta_i\\
\beta_j &~ \gamma_{i j}
\end{pmatrix}\, ,
\label{eq:g_3+1}
\end{equation}
where the Latin indices runs from $1$ to $3$. A generic 4D tensor is projected on the slice as 
\begin{equation}
{}^{(3)}T^{\mu_{\rm 1} \cdots \mu_{\rm m}}_{ \ \ \nu_{\rm 1} \cdots \nu_{\rm n}}={}^{(4)} T^{\rho_{\rm 1} \cdots \rho_{\rm m}}_{ \ \ \sigma_{\rm 1} \cdots \sigma_{\rm n}} \gamma\UD{\mu_{\rm 1}}{\rho_{\rm 1}}\cdots\gamma\UD{\mu_{\rm m}}{\rho_{\rm m}} \gamma\UD{\sigma_{\rm 1}}{\nu_{\rm 1}}\cdots\gamma\UD{\sigma_{\rm n}}{\nu_{\rm n}}\, .
\label{eq:3Dproj}
\end{equation}  

{\tt BiGONLight} is designed to accept as input directly the ADM quantities ($\alpha, \, \beta^i,\, \gamma_{i j},\, K_{i j}$) generated by a numerical simulation. However, the powerful symbolic algebra manipulation of the {\tt Wolfram} language allows also to use the analytical form of the metric $g_{\mu \nu}$ as input. 
In this case, the ADM quantities are computed in {\tt BiGONLight} by the function {\tt ADM[]} accordingly to Eqs.~\eqref{eq:K_def},~\eqref{eq:nu_nd},~\eqref{eq:g_3+1}.
On top of that, \texttt{BiGONLight} contains other functions to calculate the Christoffel symbols, \texttt{Christoffel[]}, and the Riemann tensor, \texttt{Riemann[]}. 
Note that in both cases, the input metric is provided in form of components within a specific coordinate system and not in full tensorial form. The package is designed to work in any gauge and any coordinate system, leaving the choice to the user. 
Once we have summarized the basics, we will now describe in more details and step by step the procedure for obtaining the BGO in the 3+1 decomposition: we report the 3+1 version of the geodesic equation following \cite{Vincent2012kn} and we derive all the 3+1 ingredients for the evolution equation of the BGO.

\subsection{Geodesic equation in $3+1$ decomposition}
The first code solving the $3+1$ geodesic equation was presented in \cite{hughes1994finding} and it was used to map the event horizon in numerical simulations with black holes (like heads-on collision of two black holes), while a more recent formulation of the $3+1$ geodesic equation was presented in~\cite{Vincent2012kn}. Here, we briefly resume the procedure in~\cite{Vincent2012kn}, since we will use their approach throughout all our calculations.

The null geodesics representing light rays connecting source and observer is obtained by solving 
\begin{equation}
\ell^{\sigma}\nabla_{\sigma} \ell^{\mu}=0
\label{eq:covar_geo}
\end{equation}
where $\ell^{\mu}$ is the tangent vector which obeys to the null condition
\begin{equation}
\ell^{\sigma}\ell_{\sigma}=0\, .
\label{eq:null_cond}
\end{equation}
It is $3+1$ decomposed as:
\begin{equation}
\ell^{\mu}=\calE(n^{\mu}+V^{\mu})\, ,
\label{eq:l_split}
\end{equation} 
where $\calE$ is defined by $\calE=-n^{\mu} \ell_{\mu}$ and $V^{\mu}n_{\mu}=0$. In other words, $\calE V^{\mu}$ is the component of $\ell^{\mu}$ tangent to $\Sigma_t$ and $\calE n_{\nu}$ is the orthogonal one. Substituting Eq.~\eqref{eq:l_split} in the geodesic equation~\eqref{eq:covar_geo}
after a long but straightforward calculation, see~\cite{Vincent2012kn}, the geodesic equation decouples in two differential equations
\begin{equation}
\dfrac{d \calE}{d t}= \calE \left(\alpha K_{i k} V^j V^k - V^j \partial_j \alpha \right)\, ,
\label{eq:3+1_geodE}
\end{equation}
\begin{equation}
\left\{ \begin{array}{l}
\dfrac{d x^i}{d t}= \alpha V^i - \beta^i \\
 \\
\dfrac{d V^i}{d t}= \alpha V^j \left[V^i \left( \partial_j \log \alpha - K_{j k} V^k \right) +2K^i_{\ \ j} - {}^{(3)}\Gamma^i_{\ j k} V^k \right] - \gamma^{i j} \partial_j \alpha - V^j \partial_j \beta^i
\end{array}\right.
\label{eq:3+1_geodV}
\end{equation}
for the orthogonal and the tangent components, respectively.

In the \texttt{BiGONLight} package the two functions  \texttt{EnergyEquations[]} and \texttt{GeodesicEquations[]} give the differential equations~\eqref{eq:3+1_geodE} and~\eqref{eq:3+1_geodV}, respectively. Next, we need to specify the initial conditions. The function \texttt{InitialConditions[]} is specifically constructed to get the initial conditions for $V^i$ and $\calE$ consistent with the null condition~\eqref{eq:null_cond} and the decomposition~\eqref{eq:l_split}.
The ODE~\eqref{eq:3+1_geodE} and~\eqref{eq:3+1_geodV} are then solved by using two customized versions of the Mathematica \texttt{NDSolve[]} function: \texttt{SolveEnergy[]} and \texttt{SolveGeodesic[]}.


\subsection{Parallel transport equation in the $3+1$ decomposition}
The BGO map the changes of the deviations ($\delta x^{\mu}$, $\Delta \ell^{\mu}$) between the observer $\calO$ and the source $\cal S$ along the photon geodesic $\gamma$, see Eq.~\eqref{eq:compact_W(S,O)}. This is possible only if we introduce a frame parallel transported along $\gamma$, which allows us to compare quantities at the observation point and at source position. For our purposes, the definition and the parallel transport of the frame have to be split into the 3+1 form. 
Let us start with the parallel transport of a generic vector $e^{\mu}$ along a given geodesic with tangent $\ell^{\mu}$, which is governed by
\begin{equation}
\ell^{\alpha}\nabla_{\alpha}e^{\mu}=0\, .
\label{eq:parallel_transport}
\end{equation}
The 3+1 decomposition of $e^{\mu}$ reads
\begin{equation}
e^{\mu}=C n^{\mu}+E^{\mu}\, ,
\label{eq:e_split}
\end{equation}
where we have defined the orthogonal component $C n^{\mu}=-n_{\alpha}e^{\alpha} n^{\mu}$ and the tangent component $E^{\mu}=\gamma\UD{\mu}{\alpha} e^{\alpha}$. 
The parallel transport equation~\eqref{eq:parallel_transport} becomes
\begin{equation}
n^{\mu}(n^{\alpha}\nabla_{\alpha} C+V^{\alpha}\nabla_{\alpha}C)+C(n^{\alpha}\nabla_{\alpha}n^{\mu}+V^{\alpha}\nabla_{\alpha}n^{\mu})+n^{\alpha}\nabla_{\alpha}E^{\mu}+V^{\alpha}\nabla_{\alpha}E^{\mu}=0\, .
\label{eq:PT_interm_1}
\end{equation}
Now, we make use of some $3+1$ well-known relations: for the first term we have $n^{\mu}\nabla_{\mu}C=\frac{1}{\alpha} \mathcal{L}_{\alpha \vec{n}}C$, in the second bracket we substitute $n^{\alpha}\nabla_{\alpha}n^{\mu}=D^{\mu}\log\alpha$ and the definition of the extrinsic curvature $\nabla_{\alpha}n^{\mu}=-K\UD{\mu}{\alpha}$, and for the expansion of the last two terms we use the two identities:
\begin{align*}
V^{\alpha}\nabla_{\alpha}E^{\mu}&=V^{\alpha}D_{\alpha}E^{\mu}-K_{\alpha \beta}V^{\alpha}E^{\beta} n^{\mu}\\
 \\
n^{\nu}\nabla_{\nu}E^{\mu}&=\dfrac{1}{\alpha}\left( \mathcal{L}_{\alpha \vec{n}}E^{\mu}+E^{\nu}\nabla_{\nu}(\alpha n^{\mu})\right)=\dfrac{1}{\alpha} \mathcal{L}_{\alpha \vec{n}}E^{\mu} - E^{\nu}K\UD{\mu}{\nu}+E^{\nu}\partial_{\nu}(\log\alpha) \  n^{\mu}\, .
\end{align*}
We then re-write Eq.~\eqref{eq:PT_interm_1} split into two parts, the one proportional to $n^{\mu}$ and the other tangent to $\Sigma_t$ as
\begin{equation}
\left\{\begin{array}{l}
\dfrac{1}{\alpha} \mathcal{L}_{\alpha \vec{n}}C
+V^{\nu}\partial_{\nu}C + E^{\nu}\partial_{\nu}(\log\alpha) -K_{\nu \rho}V^{\nu}E^{\rho}=0\\
 \\
\dfrac{1}{\alpha} \mathcal{L}_{\alpha \vec{n}}E^{\mu} - E^{\nu}K\UD{\mu}{\nu} +V^{\nu}D_{\nu}E^{\mu}+ C\left(\gamma^{\mu \nu}D_{\nu}\log\alpha -K\UD{\mu}{\nu}V^{\nu}\right)  =0\, ,
\end{array}\right.
\label{eq:separated_PT}
\end{equation}
where both must vanish individually in order to satisfy the parallel transport condition~\eqref{eq:parallel_transport}. The last steps consist in expanding the Lie derivative
\begin{equation}
\mathcal{L}_{\alpha \vec{n}}=\dfrac{\partial}{\partial t}-\mathcal{L}_{ \vec{\beta}}
\end{equation}
and converting partial derivatives with respect to the time $t$ into total derivative via
\begin{equation}
\dfrac{\partial}{\partial t}=\dfrac{d }{d t}-\alpha V^j\partial_j + \beta ^j\partial_j \, .
\end{equation}
The final result for the $3+1$ parallel transport equation is
\begin{equation}
\left\{ \begin{array}{l}
\dfrac{1}{\alpha} \dfrac{d C}{d t} + E^{i}\partial_{i}\log\alpha -K_{i j}V^{i}E^{j}=0\\
\\
 \dfrac{1}{\alpha}\left(\dfrac{d E^{i}}{d t}+ E^j\partial_j \beta^i \right)+ \ ^{(3)}\Gamma^i_{\ j k} V^j E^{k} - K\UD{i}{ j}E^{j} +C\left( \gamma^{i j}D_{j}\log\alpha -K\UD{i}{j}V^{j}\right)  =0\, ,
\end{array} \right.
\label{eq:3+1PT}
\end{equation}
where we have only spatial indices $i, j=1,2,3$ since all the quantities lie on $\Sigma_t$.

The system of equations~\eqref{eq:3+1PT} has to be solved for each vector of the frame we choose. 
Following \cite{grasso2019BGO}, we choose the semi-null frame (SNF) composed by the tetrad of vectors 
\begin{equation}
\phi\UD{\mu}{\boldsymbol{\alpha}}=(u^{\mu},\phi\UD{\mu}{\mathbf{1}},\phi\UD{\mu}{\mathbf{2}},\ell^{\mu})\, ,
\end{equation}
where $u^{\mu}$ is the matter four-velocity, $\ell^{\mu}$ is the tangent of the photon geodesic. The two vectors $\phi\UD{\mu}{\boldsymbol{A}}$ are  orthonormal to both $u^{\mu}$ and $\ell^{\mu}$, namely:
\begin{equation}
\left\{ \begin{array}{l}
g_{\mu \nu} u^{\mu}\phi\UD{\mu}{\mathbf{A}}=0\\
g_{\mu \nu} \ell^{\mu}\phi\UD{\mu}{\mathbf{A}}=0\\
g_{\mu \nu} u^{\mu}\ell^{\nu}=Q\\
g_{\mu \nu} \phi\UD{\mu}{\mathbf{A}}\phi\UD{\mu}{\mathbf{B}}=\delta_{\mathbf{A} \mathbf{B}}\\
\end{array}\right.
\label{eq:SNF_def_rel}
\end{equation}
with $Q$ a real number. Each vector of the SNF is then decomposed in $3+1$ form as
\begin{equation}
\phi^{\mu}_{\boldsymbol{\alpha}}=\Phi_{\boldsymbol{\alpha}}n^{\mu}+F^{\mu}_{\boldsymbol{\alpha}}\,
\label{eq:3+1vec}
\end{equation}
and parallelly propagated by solving Eq.~\eqref{eq:3+1PT} for its orthogonal and tangent components, i.e. $\Phi_{\boldsymbol{\alpha}} n^{\mu}=-n_{\sigma}\phi^{\sigma}_{\boldsymbol{\alpha}} n^{\mu}$ and $F^{\mu}_{\boldsymbol{\alpha}}=\gamma\UD{\mu}{\sigma} \phi^{\sigma}_{\boldsymbol{\alpha}}$. 

In {\tt BiGONLight} this is demanded to the function \texttt{PTransportedFrame[]}, which gives as output the components in~\eqref{eq:3+1vec} of the SNF parallel transported. The function \texttt{PTransportedFrame[]} uses \texttt{ParallelTransport[]} to obtain Eq.~\eqref{eq:3+1PT} for each vector of the frame and then solve them with a customised {\tt NDSolve[]} function.

\subsection{The optical tidal matrix and the GDE for the BGO}
\label{sec:W3+1}
So far the $3+1$ equations Eqs.~\eqref{eq:3+1_geodE}-\eqref{eq:3+1_geodV} and Eq.~\eqref{eq:3+1PT} we presented are valid for any type of geodesics and to parallelly transport any type of vectors along that geodesic. From now on, we will restrict the BGO formalism to the case of a SNF parallelly propagated along a null-like geodesic.
The equation for the BGO has then to be projected onto the SNF. The result is simply given by
\begin{equation}
\dfrac{d}{d \lambda} \mathcal{W}=\begin{pmatrix}
0 & \mathbb{1}\\
R\UDDD{\boldsymbol{\mu}}{\ell}{\ell}{\boldsymbol{\nu}} & 0
\end{pmatrix} \mathcal{W}\, ,
\label{eq:GDE_for_BGO_inframe}
\end{equation}
where the only formal difference with respect to Eq.~\eqref{eq:GDE_for_BGO} is that the covariant derivative along the photon geodesic $D/D\lambda$ reduces to the total derivative $d/d\lambda$ in the SNF.
In Eq.~\eqref{eq:GDE_for_BGO_inframe}, the optical tidal matrix is projected in the SNF and it is given by 
\beq
R\UDDD{\boldsymbol{\mu}}{\ell}{\ell}{\boldsymbol{\nu}}=h^{\boldsymbol{\mu} \boldsymbol{\omega}}R_{\boldsymbol{\omega} \ell \ell \boldsymbol{\nu}}=\phi^{\rho \boldsymbol{\mu}}R_{\rho \alpha \beta \sigma} \ell^{\alpha}\ell^{\beta} \phi\UD{\sigma}{\boldsymbol{\nu}}
\label{eq:RinSNF}
\eeq
where the inverse of the induced metric of the frame
\begin{equation}
 h^{\boldsymbol{\mu} \boldsymbol{\omega}}=\begin{pmatrix}
0 & 0 & 0 & \frac{1}{Q} \\
0 & 1 & 0 & 0 \\
0 & 0 & 1 & 0 \\
\frac{1}{Q} & 0 & 0 & \frac{1}{Q^2}
\end{pmatrix}\, 
\label{eq:frame_matrix}
\end{equation}
is used to raise the indices.
In general $R\UDDD{\boldsymbol{\mu}}{\ell}{\ell}{\boldsymbol{\nu}}$ is a $4 \times 4$ matrix with non trivial components. However, in the SNF it is easy to use the symmetries of the Riemann tensor to show that the components $R\UDDD{\boldsymbol{0}}{\ell}{\ell}{\boldsymbol{\nu}}=R\UDDD{\ell}{\ell}{\ell}{\boldsymbol{\nu}}$ and $R\UDDD{\boldsymbol{\mu}}{\ell}{\ell}{\boldsymbol{0}}=R\UDDD{\boldsymbol{\mu}}{\ell}{\ell}{\ell}$ vanish.
Now, let us use Eq.~\eqref{eq:l_split} and Eq.~\eqref{eq:3+1vec} in Eq.~\eqref{eq:RinSNF} to write the optical tidal matrix in terms of $3+1$ quantities
\beq
\begin{array}{l}
R\UDDD{\boldsymbol{\mu}}{\ell}{\ell}{\boldsymbol{\nu}}=(\Phi^{\boldsymbol{\mu}} n^{\rho} + F^{\boldsymbol{\mu} \rho})R_{\rho \alpha \beta \sigma}\mathcal{E}^2( n^{\alpha}n^{\beta}+n^{\alpha}V^{\beta}+V^{\alpha}n^{\beta}+V^{\alpha}V^{\beta})(\Phi_{\boldsymbol{\nu}} n^{\sigma} + F\UD{\sigma}{\boldsymbol{\nu}})
\end{array}
\eeq
After some tedious but straightforward calculations, we finally obtain
\begin{equation}
\begin{array}{lc}
 R\UDDD{\boldsymbol{\mu}}{\ell}{\ell}{\boldsymbol{\nu}}= & \calE^2 h^{\boldsymbol{\mu} \boldsymbol{\rho}} \left[ \mathcal{R}_{\beta \alpha}\left(\Phi_{\boldsymbol{\rho}} F\UD{\beta}{\boldsymbol{\nu}}V^{\alpha}+\Phi_{\boldsymbol{\nu}} F\UD{\beta}{\boldsymbol{\rho}}V^{\alpha}-\Phi_{\boldsymbol{\rho}}\Phi_{\boldsymbol{\nu}} V^{\beta}V^{\alpha}-F\UD{\alpha}{\boldsymbol{\rho}}F\UD{\beta}{\boldsymbol{\nu}}\right)+\right.\\
 & \mathcal{C}_{\sigma \beta \alpha}\left(\Phi_{\boldsymbol{\rho}} F\UD{\sigma}{\boldsymbol{\nu}}V^{\alpha}V^{\beta}+\Phi_{\boldsymbol{\nu}} F\UD{\sigma}{\boldsymbol{\rho}}V^{\alpha}V^{\beta}-F\UD{\alpha}{\boldsymbol{\rho}} F\UD{\sigma}{\boldsymbol{\nu}}V^{\beta}-F\UD{\alpha}{\boldsymbol{\nu}} F\UD{\sigma}{\boldsymbol{\rho}}V^{\beta} \right)+\\
 & \left.\mathcal{G}_{\omega \alpha \beta \sigma} F\UD{\omega}{\boldsymbol{\rho}}V^{\alpha}V^{\beta} F\UD{\sigma}{\boldsymbol{\nu}}\right]
\end{array}
\label{eq:3+1_opt}
\end{equation}
where
\begin{align}
 \mathcal{G}_{\mu \alpha \beta \nu}&=R_{\rho \delta \theta \sigma}\gamma\UD{\rho}{\mu}\gamma\UD{\delta}{\alpha}\gamma\UD{\theta}{\beta}\gamma\UD{\sigma}{\nu}= {}^{(3)}R_{\mu \alpha \beta \nu}+ K_{\mu \beta} K_{\alpha \nu}-K_{\mu \nu} K_{\beta \alpha} \label{eq:Gauss}\\
 \mathcal{C}_{\mu \alpha \beta}&=R_{\rho \delta \theta \sigma}n^{\rho}\gamma\UD{\delta}{\alpha}\gamma\UD{\theta}{\beta}\gamma\UD{\sigma}{\mu}= D_{\alpha}K_{\mu \beta}-D_{\mu}K_{\alpha \beta} \label{eq:Codazzi}\\
 \mathcal{R}_{\mu \nu}&=R_{\rho \delta \theta \sigma}n^{\rho}\gamma\UD{\delta}{\nu}\gamma\UD{\theta}{\mu}n^{\sigma}= \mathcal{L}_{\mathbf{n}}K_{\nu \alpha}+\dfrac{1}{\alpha}D_{\nu}D_{\alpha}\alpha+ K\UD{\rho}{\alpha}K_{\nu \rho}
\label{eq:Ricci}
\end{align}
are the Gauss relation, the Codazzi relation and the Ricci relation, respectively (see e.g. \cite{baumgarte2010numerical}).
In \texttt{BiGONLight}, the three functions \texttt{GaussRelation[]}, \texttt{CodazziRelation[]}, \texttt{RicciRelation[]} compute the relations in Eqs.~\eqref{eq:Gauss}-\eqref{eq:Ricci} and the function \texttt{OpticalTidalMatrix[]} collects all previous results together in order to obtain the optical tidal matrix $R\UDDD{\boldsymbol{\mu}}{\ell}{\ell}{ \boldsymbol{\nu}} / \calE^2$ in Eq.~\eqref{eq:3+1_opt} as output.
We now have the equation for the BGO, Eq.~\eqref{eq:GDE_for_BGO_inframe}, projected onto the SNF and written in terms of $3+1$ quantities. To solve it, it is easier to change the derivation variable from the affine parameter $\lambda$ to the time $t$ according to
\begin{equation}
\dfrac{d }{d \lambda}=\dfrac{d \, t}{d \lambda}\dfrac{d }{d t}=\frac{ \mathcal{E}}{\alpha}\dfrac{d }{d t} \, ,
\end{equation}
where we used the time component of the tangent vector in Eq.~\eqref{eq:l_split}, i.e. $\ell^0=d \, t / d \lambda= \mathcal{E} / \alpha$.
The GDE for the BGO, Eq.~\eqref{eq:GDE_for_BGO_inframe}, decouples in two systems of first-order ODE, one for $(\WXX, \WLX)$ and the other for $(\WXL, \WLL)$, computed separately using the function \texttt{BGOequations[]}:
\begin{equation}
\left\{\begin{array}{l}
\dfrac{d \, \WXX{}\UD{\boldsymbol{\mu}}{\boldsymbol{\nu}}}{d t}=\dfrac{\alpha}{ \mathcal{E}}\, \WLX{}\UD{\boldsymbol{\mu}}{\boldsymbol{\nu}}\\
 \\
\dfrac{d \, \WLX{}\UD{\boldsymbol{\mu}}{\boldsymbol{\nu}}}{d t}=\dfrac{\alpha}{ \mathcal{E}}\, R\UDDD{\boldsymbol{\mu}}{\ell}{\ell}{\boldsymbol{\sigma}}\WXX{}\UD{\boldsymbol{\sigma}}{\boldsymbol{\nu}}
\end{array}\right.\ , \
\left\{\begin{array}{l}
\dfrac{d \, \WXL{}\UD{\boldsymbol{\mu}}{\boldsymbol{\nu}}}{d t}=\dfrac{\alpha}{ \mathcal{E}}\, \WLL{}\UD{\boldsymbol{\mu}}{\boldsymbol{\nu}}\\
 \\
\dfrac{d \, \WLL{}\UD{\boldsymbol{\mu}}{\boldsymbol{\nu}}}{d t}=\dfrac{\alpha}{ \mathcal{E}}\, R\UDDD{\boldsymbol{\mu}}{\ell}{\ell}{\boldsymbol{\sigma}}\WXL{}\UD{\boldsymbol{\sigma}}{\boldsymbol{\nu}}
\end{array}\right.
\label{eq:GDE_BGO_syst}
\end{equation}
with initial conditions:
\begin{equation}
\left\{\begin{array}{l}
\WXX{}\UD{\boldsymbol{\mu}}{\boldsymbol{\nu}}\left|_{\rm \calO}\right.=\delta\UD{\boldsymbol{\mu}}{\boldsymbol{\nu}}\\
\WXL{}\UD{\boldsymbol{\mu}}{\boldsymbol{\nu}}\left|_{\rm \calO}\right.=0\\
\WLX{}\UD{\boldsymbol{\mu}}{\boldsymbol{\nu}}\left|_{\rm \calO}\right.=0\\
\WLL{}\UD{\boldsymbol{\mu}}{\boldsymbol{\nu}}\left|_{\rm \calO}\right.=\delta\UD{\boldsymbol{\mu}}{\boldsymbol{\nu}}\\
\end{array}\right.
\label{eq:GDE_BGO_IC}
\end{equation}
The systems in Eq.~\eqref{eq:GDE_BGO_syst} are solved separately in {\tt BiGONLight} by \texttt{SolveBGO[]}. Getting the BGO with the procedure just described is one important part of {\tt BiGONLight}. Let us recall that the only inputs required are the spacetime metric, the observer four-velocity components and the initial and ending points.

With initial conditions in Eq.~\eqref{eq:GDE_BGO_IC}, Eq.~\eqref{eq:GDE_BGO_syst} gives the BGO $\mathcal{W}(p_\lambda,\mathcal{O})$ integrated backward in time from the observer. The relation with the BGO $\mathcal{W}(p_\lambda,\mathcal{S})$ integrated forward in time from the source can be explicitly written down for all the BGO components from Eq.~\eqref{eq:W(p,o)_from_W(p,s)}. To this end we need the inverse matrix $\mathcal{W}^{-1}$ which is easily found from the symplectic property 
\begin{equation}
\calW^T{}\UD{\tilde{m}}{\tilde{a}} \Omega_{\tilde{m} \tilde{s}} \calW\UD{\tilde{s}}{\tilde{b}} =\Omega_{\tilde{a} \tilde{b}}\, ,
\label{eq:propW_symplectic}
\end{equation}
where $\Omega$ is the $8 \times 8$ non-singular, skew-symmetric matrix 
\begin{equation}
\Omega_{\tilde{a} \tilde{b}}=\begin{pmatrix}
0 & h_{\boldsymbol{\alpha} \boldsymbol{\beta}}\\
-h_{\boldsymbol{\gamma} \boldsymbol{\delta}} & 0
\end{pmatrix}\, ,
\end{equation}
with Latin tilded indices running from $0$ to $7$ ($\tilde{a}, \dots=0,1,\dots,7 $) and the Greek bold indices ($\boldsymbol{\alpha}=0,1,\dots,3$) indicate the components in the SNF. 
By inverting Eq.~\eqref{eq:propW_symplectic} we find
\begin{align}
\calW^{-1}&=\Omega^{-1 }\calW^T \Omega \nonumber\\
&=\begin{pmatrix}
0 & -h^{\boldsymbol{\alpha} \boldsymbol{\rho}}\\
h^{\boldsymbol{\beta} \boldsymbol{\sigma}} & 0
\end{pmatrix}\begin{pmatrix}
\WXX{}\UD{\boldsymbol{\nu}}{\boldsymbol{\sigma}} & \WLX{}\UD{\boldsymbol{\mu}}{\boldsymbol{\sigma}}\\
\WXL{}\UD{\boldsymbol{\nu}}{\boldsymbol{\rho}} & \WLL{}\UD{\boldsymbol{\mu}}{\boldsymbol{\rho}}
\end{pmatrix}\begin{pmatrix}
0 & h_{\boldsymbol{\nu} \boldsymbol{\gamma}}\\
-h_{\boldsymbol{\mu} \boldsymbol{\delta}} & 0
\end{pmatrix}\, .
\label{eq:derive_W_inverse}
\end{align}
The transformation from forward to backward BGO in Eq.~\eqref{eq:W(p,o)_from_W(p,s)} finally reads
\begin{align}
\WXX(p_{\lambda}, \calO){}\UD{\boldsymbol{\sigma}}{\boldsymbol{\nu}}&=  \WXX(p_{\lambda}, \calS){}\UD{\boldsymbol{\sigma}}{\boldsymbol{\alpha}} h^{\boldsymbol{\alpha} \boldsymbol{\rho}} \, \WLL^T( \calO,\calS){}\UD{\boldsymbol{\mu}}{\boldsymbol{\rho}} \, h_{\boldsymbol{\mu} \boldsymbol{\nu}}+ \nonumber\\
&-\WXL(p_{\lambda}, \calS){}\UD{\boldsymbol{\sigma}}{\boldsymbol{\alpha}}h^{\boldsymbol{\alpha} \boldsymbol{\rho}} \, \WLX^T( \calO,\calS){}\UD{\boldsymbol{\mu}}{\boldsymbol{\rho}} \, h_{\boldsymbol{\mu} \boldsymbol{\nu}}\label{eq:WXX_inverse}\\
\WXL(p_{\lambda}, \calO){}\UD{\boldsymbol{\sigma}}{\boldsymbol{\nu}}&= - \WXX(p_{\lambda}, \calS){}\UD{\boldsymbol{\sigma}}{\boldsymbol{\alpha}} h^{\boldsymbol{\alpha} \boldsymbol{\rho}} \, \WXL^T( \calO,\calS){}\UD{\boldsymbol{\mu}}{\boldsymbol{\rho}} \, h_{\boldsymbol{\mu} \boldsymbol{\nu}}+ \nonumber\\
&+\WXL(p_{\lambda}, \calS){}\UD{\boldsymbol{\sigma}}{\boldsymbol{\alpha}}h^{\boldsymbol{\alpha} \boldsymbol{\rho}} \, \WXX^T( \calO,\calS){}\UD{\boldsymbol{\mu}}{\boldsymbol{\rho}} \, h_{\boldsymbol{\mu} \boldsymbol{\nu}}\label{eq:WXL_inverse}\\
\WLX(p_{\lambda}, \calO){}\UD{\boldsymbol{\sigma}}{\boldsymbol{\nu}}&=  \WLX(p_{\lambda}, \calS){}\UD{\boldsymbol{\sigma}}{\boldsymbol{\alpha}} h^{\boldsymbol{\alpha} \boldsymbol{\rho}} \, \WLL^T( \calO,\calS){}\UD{\boldsymbol{\mu}}{\boldsymbol{\rho}} \, h_{\boldsymbol{\mu} \boldsymbol{\nu}}+ \nonumber\\
&-\WLL(p_{\lambda}, \calS){}\UD{\boldsymbol{\sigma}}{\boldsymbol{\alpha}}h^{\boldsymbol{\alpha} \boldsymbol{\rho}} \, \WLX^T( \calO,\calS){}\UD{\boldsymbol{\mu}}{\boldsymbol{\rho}} \, h_{\boldsymbol{\mu} \boldsymbol{\nu}}\label{eq:WLX_inverse}\\
\WLL(p_{\lambda}, \calO){}\UD{\boldsymbol{\sigma}}{\boldsymbol{\nu}}&= - \WLX(p_{\lambda}, \calS){}\UD{\boldsymbol{\sigma}}{\boldsymbol{\alpha}} h^{\boldsymbol{\alpha} \boldsymbol{\rho}} \, \WXL^T( \calO,\calS){}\UD{\boldsymbol{\mu}}{\boldsymbol{\rho}} \, h_{\boldsymbol{\mu} \boldsymbol{\nu}}+ \nonumber\\
&+\WLL(p_{\lambda}, \calS){}\UD{\boldsymbol{\sigma}}{\boldsymbol{\alpha}}h^{\boldsymbol{\alpha} \boldsymbol{\rho}} \, \WXX^T( \calO,\calS){}\UD{\boldsymbol{\mu}}{\boldsymbol{\rho}} \, h_{\boldsymbol{\mu} \boldsymbol{\nu}}
\label{eq:WLL_inverse}
\end{align}
These relations are coded in the section ``Forward to backward transformation for $\mathcal{W}$ operators'' of each sample notebook in the repository {\color{blue}{\tt {https://github.com/MicGrasso/\\ 
bigonlight.git}}}.


\section{Optical observables with {\tt BiGONLight}}
\label{sec:obs}
The recipe to obtain optical observables using \texttt{BiGONLight} can be summarized as follows. One needs to:
\begin{enumerate}
\item specify the spacetime metric $g_{\mu \nu}$ and the source $\mathcal{S}$ and observer $\mathcal{O}$ kinematics, namely four-velocity $u^{\mu}$ and four-acceleration $w^{\mu}$. They can be given already in $3+1$ components or as 4D quantities and the functions {\tt ADM[]} and {\tt Vsplit[]} will do the splitting of $g_{\mu \nu}$, and $u^{\mu}$ and $w^{\mu}$ respectively;
\item set up the initial photon geodesic using {\tt Vsplit[]} for the null tangent $\ell^{\mu}$, which gives its $3+1$ components $\calE$ and $V^{i}$. Alternatively one can give $\calE$ and a vector $V^{i}$ that has to be assigned by specifying the spatial direction $V^2$ and $V^3$ and use {\tt InitialConditions[]} to get $V^1$ from the null condition; \\
\item obtain the geodesic equations Eq.~\eqref{eq:3+1_geodE} and Eq.~\eqref{eq:3+1_geodV} from {\tt GeodesicEquations[]} and {\tt EnergyEquations[]}, and then solve them with {\tt SolveGeodesic[]} and {\tt SolveEnergy[]}\footnote{For the purposes of our paper, we need to solve only photon geodesics. However, the code can be used to trace any type of geodesics, namely time-like and space-like also, by specifying the appropriate initial tangent vector in the $3+1$ splitting with {\tt Vsplit[]}.};\\
\item set up the initial conditions for the SNF according to Eq.~\eqref{eq:SNF_def_rel}, directly in $3+1$ components or using {\tt SNF[]}, which is specifically designed to compute the SNF in $3+1$. Then {\tt PTransportedFrame[]} will give the SNF parallel transported along the light ray;\\
\item compute separately $R\UDDD{\boldsymbol{\mu}}{\ell}{\ell}{\boldsymbol{\nu}}$ projected into the SNF with {\tt OpticalTidalMatrix[]};\\
\item obtain the ODE system for the BGO in Eq.~\eqref{eq:GDE_BGO_syst} with {\tt BGOequations[]} and, together with the initial conditions in Eq.~\eqref{eq:GDE_BGO_IC}, solve it using {\tt SolveBGO[]} to finally find the full $\mathcal{W}$ matrix;\\
\end{enumerate}
Note that the components of the optical tidal matrix can have a very complicated expression, that may cause problems in solving the GDE \eqref{eq:GDE_BGO_syst}. This can be overcame by using an interpolated form of $R\UDDD{\boldsymbol{\mu}}{\ell}{\ell}{\boldsymbol{\nu}}$: the {\tt Mathematica Interpolation[]} function allows to use different methods and reach an excellent precision, see Sec.\ref{sec:test}.

Step (vi) is the starting point to compute the optical observables, which are all given by different combinations and/or functions of the $\mathcal{W}$ components. It is important to remark that all the observables in the BGO formalism are written in terms of $\mathcal{W}(\calS, \calO)$, namely the map computed from the observer to the source. As we already recalled, this is obtained directly by integrating the GDE backward in time. However in some cases, e.g. if the spacetime model comes from a numerical simulation, it may be more convenient to get the inverse BGO map $\mathcal{W}^{-1}(\calS, \calO)$ and then use the transformations Eqs.~\eqref{eq:WXX_inverse}-\eqref{eq:WLL_inverse} to obtain the $\mathcal{W}$ needed for the observables. 

Here we list the four observables that we study in this paper. The redshift is simply given by its definition
\begin{equation}
1+z = \dfrac{\left(\ell_{\sigma } u^{\sigma}\right)|_{\cal S}}{\left(\ell_{\sigma } u^{\sigma}\right)|_{\cal O}}\, ,
\label{eq:redshift_def}
\end{equation}
where $\ell^{\sigma}$ is the tangent to the light ray, and $u^{\sigma}_{\mathcal{O}}$ and $u^{\sigma}_{\mathcal{S}}$ are the observer and source four-velocities. The same definition in $3+1$ splitting reads:
\begin{equation}
1+z=\frac{\mathcal{E}_{\mathcal{S}}}{\mathcal{E}_{\mathcal{O}}}\,\frac{1-\left(\gamma _{ij} V^i U^j\right)|_{\mathcal{S}}}{1-\left(\gamma _{ij} V^i U^j\right)|_{\mathcal{O}}}\left[\frac{1- \left(\gamma _{ij}U^i U^j\right)|_{\mathcal{S}}}{1-\left(\gamma _{ij}U^i U^j\right)|_{\mathcal{O}}}\right]^{\frac{1}{2}}\, .
\label{eq:redshift_def_3+1}
\end{equation}
The angular diameter distance is formally given by
\begin{equation}
D_{ang} = \left(\ell_{\sigma} u^{\sigma}\right)|_{\cal O} \left| \det \left(\WXL {}\UD{\bm A}{\bm B}\right) \right|^{\frac{1}{2}}\, ,
\label{eq:D_ang_BGO}
\end{equation}
where $\WXL {}\UD{\bm A}{\bm B}$ is the map between the physical size of the source and the angle subtended in the sky, as measured at the observer position, namely
\begin{equation}
\delta \theta^{\bm A}= {\left( \ell_{\sigma} u^{\sigma}\right)|_{\cal O}}^{-1} \left(\WXL {}\UD{\bm A}{\bm B}\right)^{-1} \delta x_{\mathcal{S}}^{\bm B}\, .
\end{equation}
Conversely, the parallax distance is related to the displacement of the observer position and the apparent angular shift of the source position, as measured from the observer
\begin{equation}
\delta \theta^{\bm A}= - {\left( \ell_{\sigma} u^{\sigma}\right)|_{\cal O}}^{-1} \left(\WXL {}\UD{\bm A}{\bm C}\right)^{-1} \WXX {}\UD{\bm C}{\bm B} \delta x_{\mathcal{O}}^{\bm B}\, .
\end{equation}
The expression for the parallax distance is
\begin{equation}
D_{par} =\left(\ell_{\sigma} u^{\sigma}\right)|_{\cal O} \frac{\left| \det \left(\WXL {}\UD{\bm A}{\bm B}\right) \right|^{\frac{1}{2}}}{\left| \det \left(\WXX {}\UD{\bm A}{\bm B}\right) \right|^{\frac{1}{2}}}\, .
\label{eq:D_par_BGO}
\end{equation}
The last observable that we consider in this paper is the redshift drift $\zeta$, given in terms of the BGO by, \cite{Julius}
\begin{equation}
\zeta\equiv\frac{\delta \log(1+z)}{\delta \tau_{\mathcal{O}}}= \Xi_{\rm Doppler}- \left( u_{\mathcal{O}} , \, \frac{u_{\mathcal{S}}}{1+z} \right) \boldsymbol{U} \begin{pmatrix} u_{\mathcal{O}}\\ \dfrac{u_{\mathcal{S}}}{1+z}\end{pmatrix}\, .
\label{eq:z_DRIFT_BGO}
\end{equation}
In the above expression $\tau_{\calO}$ is the proper time of the observer, the first term
\begin{equation}
\Xi_{\rm Doppler}=\left[\dfrac{1}{1+z}\dfrac{\left(\ell^{\bm \mu}w_{\bm \mu}\right)|_{\mathcal{S}}}{\left(\ell^{\bm \mu} u_{\bm \mu}\right)|_{\mathcal{S}}}-\dfrac{\left(\ell^{\bm \mu}w_{\bm \mu}\right)|_{\mathcal{O}}}{\left(\ell^{\bm \mu} u_{\bm \mu}\right)|_{\mathcal{O}}}\right]
\label{eq:doppler}
\end{equation}
represents the Doppler effect caused by the four-acceleration $w^{\sigma}$ of the observer and the source, and $U$ is an $8 \times 8$ matrix given by the following combinations of the BGO
\begin{equation}
U=\begin{pmatrix}
- \WXL^{-1}{}\UD{\bm \nu}{\bm \rho}\WXX{}\UD{\bm \rho}{\bm \sigma} & \WXL^{-1}{}\UD{\bm \nu}{\bm \rho} \\
\WLL{}\UD{\bm \mu}{\bm \nu}\WXL^{-1}{}\UD{\bm \nu}{\bm \rho}\WXX{}\UD{\bm \rho}{\bm \sigma}-\WLX{}\UD{\bm \mu}{\bm \sigma} & - \WLL{}\UD{\bm \mu}{\bm \nu} \WXL^{-1}{}\UD{\bm \nu}{\bm \rho}
\end{pmatrix} \, .
\end{equation}
Let us notice that, even if the BGO formalism is independent of the specific choice of the frame used, the observables are dependent on this choice, as evident by the explicit dependence on $u^{\mu}_{\mathcal{O}}$, $u^{\mu}_{\mathcal{S}}$, $w^{\mu}_{\mathcal{O}}$ and $w^{\mu}_{\mathcal{S}}$ in Eqs. \eqref{eq:D_ang_BGO}-\eqref{eq:doppler}. Indeed, it is possible to transform locally between two different frames using an appropriate Lorentz transformation, but this modifies the observables introducing special relativistic effects like Doppler effect or aberration.

The reader can find the derivation of the Eqs.~\eqref{eq:D_ang_BGO}-\eqref{eq:z_DRIFT_BGO} in \cite{grasso2019BGO, korzynski2018optical, Julius}). All the expressions in Eqs.~\eqref{eq:D_ang_BGO}, \eqref{eq:D_par_BGO} and \eqref{eq:z_DRIFT_BGO} are new with respect to the standard approach, in the sense that these observables are expressed within a new, unified framework. However, while for $D_{\rm ang}$ and $D_{\rm par}$ there already exist analogous formulas, where instead of the BGO we have the magnification and the parallax matrix (see \cite{korzynski2020geometric} for the comparison), it did not exist a general formula for the redshift drift: Eq.~\eqref{eq:z_DRIFT_BGO} looks the same for every spacetime model under consideration. Instead, in the standard approach the redshift drift is calculated by taking the derivative with respect to the time coordinate of the definition of the redshift, and this depends on the specific form of the metric tensor and null-geodesic, the latter depending in turns on the symmetries that one gives to the initial conditions. This means that the equations to get the redshift drift in the standard approach look different for each specific model and/or configuration of the light rays\footnote{To be more precise, Eq.~\eqref{eq:z_drift_FLRW} is valid for the FLRW model only as well as Eq.~\eqref{eq:z_drift_Szekeres} is valid for the Szekeres model and geodesics along the symmetry axis only. Instead, Eq.~\eqref{eq:z_DRIFT_BGO} looks the same in both cases.}.


\section{Code tests}
\label{sec:test}


In this section we test the accuracy of \texttt{BiGONLight} within well-known cosmological models. The tests are performed by considering the following observables: redshift, angular diameter distance, parallax distance, and redshift drift. We compare the results obtained with three different procedures, by defining the estimator $\Delta O({\rm BGO, X})$
\begin{equation}
\Delta O({\rm BGO, X}) \equiv \dfrac{O^{\rm BGO}-O^{\rm X}}{O^{\rm X}}\, ,
\label{eq:deltaO_def}
\end{equation}
where $O^{\rm BGO}$ refers to Eqs.~\eqref{eq:redshift_def},~\eqref{eq:D_ang_BGO},~\eqref{eq:D_par_BGO}, and~\eqref{eq:z_DRIFT_BGO}. We consider the following three cases: (i) the $\Lambda$CDM model, where the specetime metric is the analytical input for {\tt BiGONLight} to compute $O^{\rm BGO}$ and for $O^{\rm X}$ we use the analytical well-known solutions for all the four observables, see Sec.~\ref{sec:LCDM}; (ii) the inhomogeneous Szekeres model \cite{Meures:Szekeres}, where the specetime metric is again the analytical input for {\tt BiGONLight} but to obtain $O^{\rm X}$ we solve numerically a specific differential equation for each observable, see Sec.~\ref{sec:Szekeres}; (iii) the Einstein-de Sitter model, where the input for {\tt BiGONLight} are the $3+1$ quantities coming from the {\tt Einstein Toolkit} simulation and $O^{\rm X}$ is obtained analytically, see Sec.~\ref{sec:ET}.

It worth noting that if $O^{\rm X}$ is obtained analytically, then $\mathrm{max}\left|\Delta O({\rm BGO, X})\right|$ represents the simulation error: in case (i) we have just the computational error from {\tt BiGONLight} whereas in case (iii) the final error in the observables is the combined effect of both the {\tt Einstein Toolkit} and {\tt BiGONLight} finite precision. On the other hand, if $O^{\rm X}$ is obtained numerically, as in case (ii), then $\Delta O({\rm BGO, X})$ gives only an estimation of the accuracy of the two methods used.

\subsection{The $\Lambda$CDM model}
\label{sec:LCDM}
The first group of tests regards the study of light propagation in the flat $\Lambda$CDM model. This is an exact solution of the Einstein field equations representing an homogeneous and isotropic spacetime and the matter-energy content consists of irrotational dust of cold dark matter and a cosmological constant $\Lambda$. The line element is given by
\begin{equation}
ds^2=a(\eta)^2\left(-d\eta^2+ \delta_{i j} dx^idx^j \right)
\label{eq:FLRW_ds}
\end{equation} 
where $\eta$ is the conformal time and $a(\eta)$ is the scale factor, which is the solution of Einstein equations and describes the dynamics of the model. The explicit result is found to be \cite{gradshteyn2014table}
\begin{equation}
a(\eta)=\frac{\sqrt[3]{\frac{\Omega_{\rm m_0}}{\Omega_{\rm \Lambda}}} \Big(1-{\rm cn}\left(\mathit{y} |\mathit{r} \right)\Big)}{(\sqrt{3}-1)+(\sqrt{3}+1) {\rm cn}\left(\mathit{y} |\mathit{r} \right)}\, ,\label{eq:a_LCDM}
\end{equation}
where ${\rm cn}(\mathit{y}|\mathit{r})$ is the Jacobi elliptic cosine function, with $\mathit{y}=\left(\sqrt[4]{3} \sqrt[6]{\Omega_{\rm \Lambda}} \sqrt[3]{\Omega_{\rm m_0}}\right) \mathcal{H}_0 \eta$, $\mathcal{H}_{\rm 0}$ being the Hubble parameter $\mathcal{H}=\frac{1}{a}\frac{d a}{d \eta}$ evaluated today, and $\mathit{r}=\sqrt{\frac{\sqrt{3}+2}{4}}$.

To test {\tt BiGONLight} we consider two classical observables, namely the redshift $z$ and the angular diameter distance $D_{\rm ang}$, and two interesting observables that are not yet measured in the cosmological context as they belong to the new research field named Real-time Cosmology, see Ref.~\cite{Quercellini:2010zr}. One is the parallax distance which exploits the motion of the Solar System with respect to the Cosmic Microwave Background frame providing a baseline of $78 \, {\rm AU}$ per year for the cosmic parallax\footnote{For an exhaustive definition of the parallax see e.g.  Ref.~\cite{rasanen2014covariant}, in which the author distinguishes between the three different cases: one source observed by two  observers separated by spacelike interval (classic parallax), two sources observed by two  observers separated by spacelike interval and two sources observed by one observer at two different moments (cosmic parallax also known as position drift).}. Cosmic parallax was first proposed in $1986$ in Ref.~\cite{kardashev} and it is expected to be measured by the Gaia satellite,~\cite{refId0}, in the next few years. For discussions and forecasts about the measurements of the cosmological parallax distance we refer to Refs.~\cite{Ding:2009xs, Quartin:2009xr, Quercellini:2008ty, Quercellini:2010zr, rasanen2014covariant, singal2015cosmological, Marcori:2018cwn, PhysRevLett.121.021101} and Refs. therein. The other is the redshift drift, i.e. the time variation of the redshift of a source. It was first derived for the FLRW models in Refs.~\cite{Sandage1962, McVittie1962}. Since then, and particularly in recent years, a lot of work has been done to investigate the measurability of the redshift drift in cosmology and the information gained, see e.g. Refs.~\cite{Corasaniti:2007bg, Uzan:2007tc, Uzan:2008qp, Martinelli:2012vq,  Lazkoz:2017fvx}.

The analytical expressions in the flat FLRW cosmologies for the four observables that we consider are
\begin{align}
z^{\rm \Lambda CDM} =& \dfrac{a_0}{a}-1 \label{eq:z_FLRW}\\
D^{\rm \Lambda CDM}_{\rm ang}=&\dfrac{a_0}{1+z}\displaystyle\int^z_0 \dfrac{d z'}{ (1+z')\mathcal{H}(z')}
 \label{eq:D_ang_FLRW}\\
\nonumber \\
D^{\rm \Lambda CDM}_{\rm par}=& \dfrac{a_0}{\mathcal{H}_{\rm 0}}\dfrac{\displaystyle\int^z_0 \dfrac{\mathcal{H}_{\rm 0} d z'}{ (1+z')\mathcal{H}(z')}}{1+\displaystyle\int^z_0 \dfrac{\mathcal{H}_{\rm 0} d z'}{ (1+z')\mathcal{H}(z')}}
 \label{eq:D_par_FLRW}\\
\nonumber  \\
\zeta^{\rm \Lambda CDM}=& \dfrac{\mathcal{H}_{\rm 0}}{a_{\rm 0}}\left(1-\dfrac{\mathcal{H}(z)}{\mathcal{H}_{\rm 0}}\right)
\, , \label{eq:z_drift_FLRW}
\end{align}
where the Hubble parameter in the $\Lambda$CDM model is given by
\begin{equation}
\mathcal{H}(z)=\mathcal{H}_{\rm 0}\sqrt{\Omega_{\rm m_0}(1+z)+\Omega_{\rm \Lambda}(1+z)^{-2}}\, .
\label{eq:E(z)}
\end{equation}
We normalize the today scale factor $a_{\rm 0}=1$ and we take $\mathcal{H}_{\rm 0}=67.36\ {\rm km\, s^{-1}\, Mpc^{-1}}$, the matter parameter today $\Omega_{\rm m_0}=0.315$, and the cosmological constant parameter  $\Omega_{\Lambda}=0.685$ from  \cite{planck2018param}.
The integral in Eq.~\eqref{eq:D_ang_FLRW} and~\eqref{eq:D_par_FLRW} can be solved analytically and the results is, \cite{gradshteyn2014table}
\begin{equation}
\displaystyle\int^z_0 \dfrac{d z'}{ (1+z')\mathcal{H}(z')}=\dfrac{
{\rm F}\big[ \xi(z) |\mathit{r} \big] - {\rm F}\big[ \xi(0)|\mathit{r} \big]}{ (\Omega_{\rm m_0})^{\frac{1}{3}}(\Omega_{\rm \Lambda})^{\frac{1}{6}} 3^{\frac{1}{4}}}
\end{equation}
where ${\rm F}\big[ \xi(z) | \mathit{r} \big]$ the elliptic integral of the first kind, with arguments $\mathit{r}=\sqrt{\frac{2+\sqrt{3}}{4}}$ and
\begin{equation}
\xi(z)=\arccos \left( \frac{2\sqrt{3}}{1+\sqrt{3}+(1+z)\sqrt[3]{\frac{\Omega_{\rm m_0}}{\Omega_{\rm \Lambda}}}}-1 \right)\, .
\end{equation}

\begin{figure}[ht]
    \centering
    \begin{subfigure}{0.49\linewidth}
        \includegraphics[width=\linewidth]{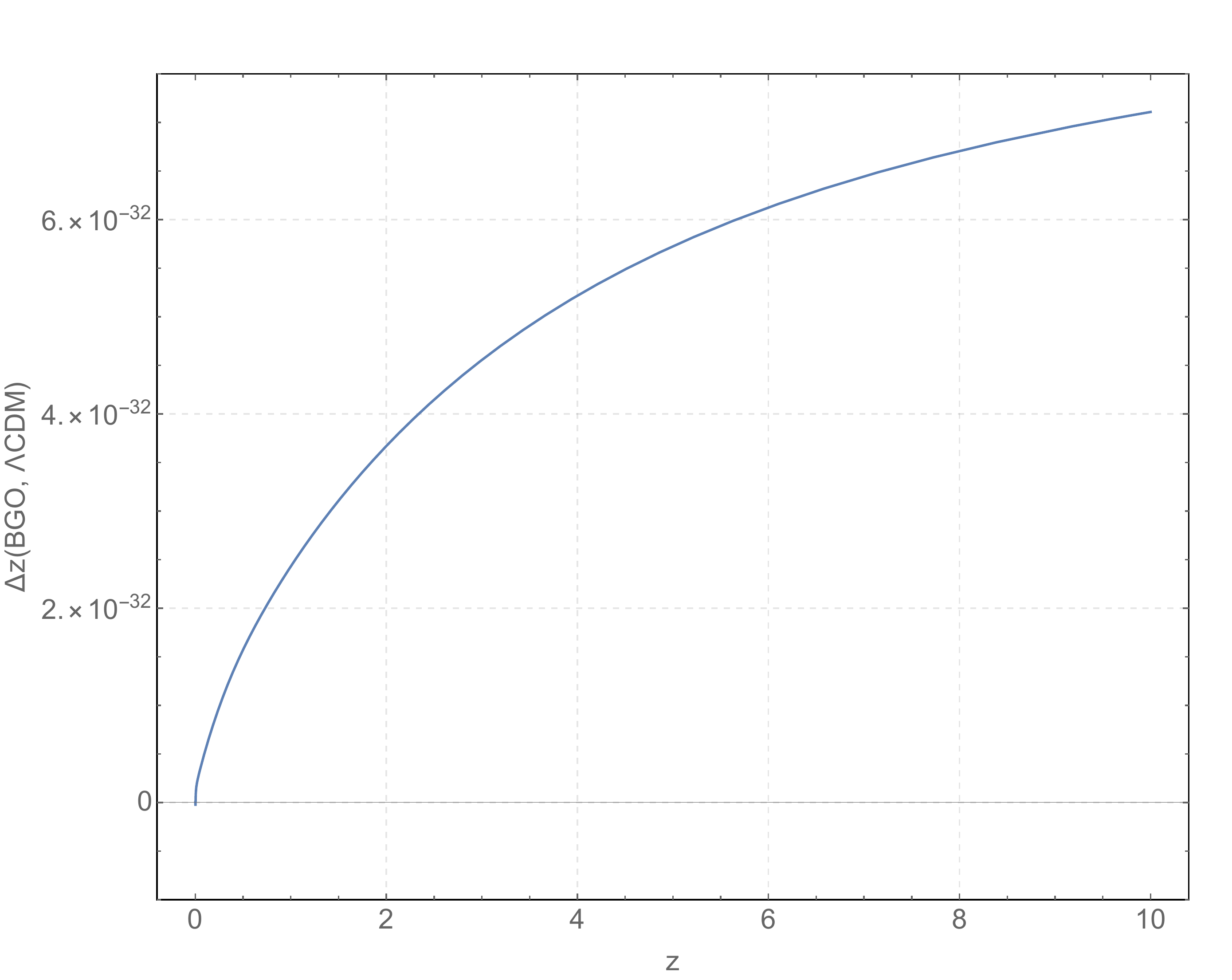}
       \caption{Redshift}
        \label{fig:LCDM_z}
    \end{subfigure}
    \begin{subfigure}{0.49\linewidth}
        \includegraphics[width=\linewidth]{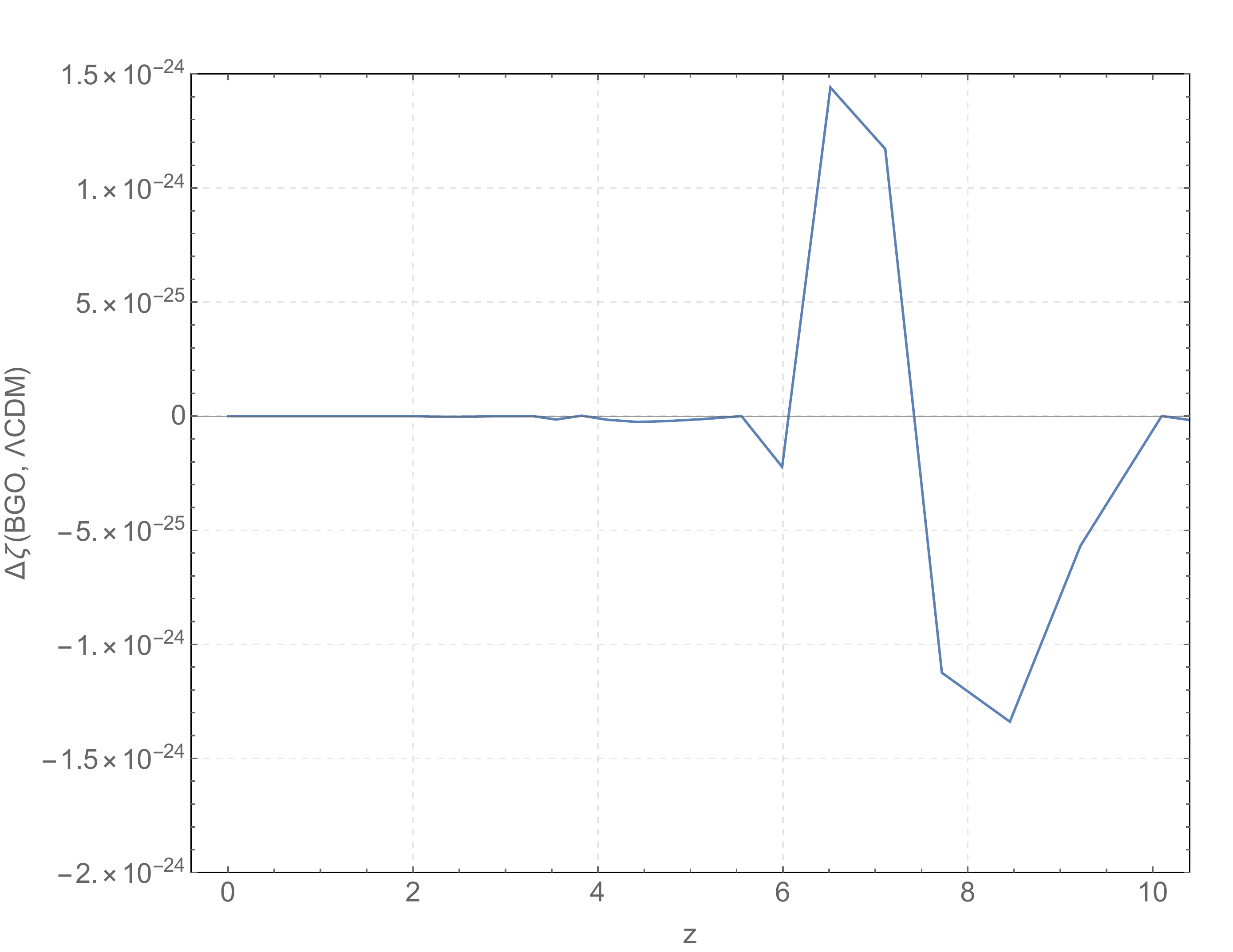}
        \caption{Redshift drift}
        \label{fig:LCDM_drift}
    \end{subfigure}
    \begin{subfigure}{0.49\linewidth}
        \includegraphics[width=\linewidth]{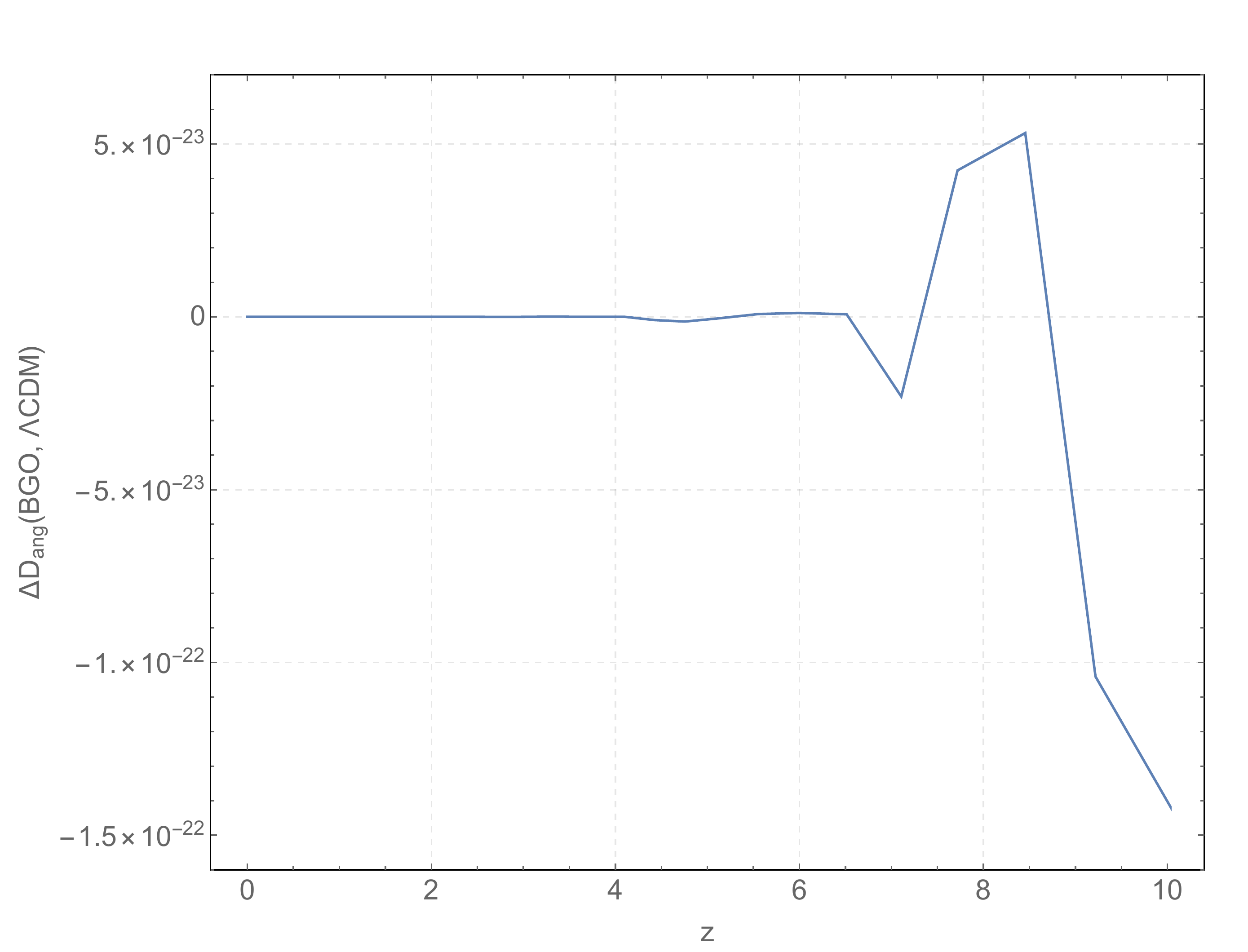}
        \caption{Angular diameter distance}
        \label{fig:LCDM_Dang}
    \end{subfigure}
    \begin{subfigure}{0.49\linewidth}
        \includegraphics[width=\linewidth]{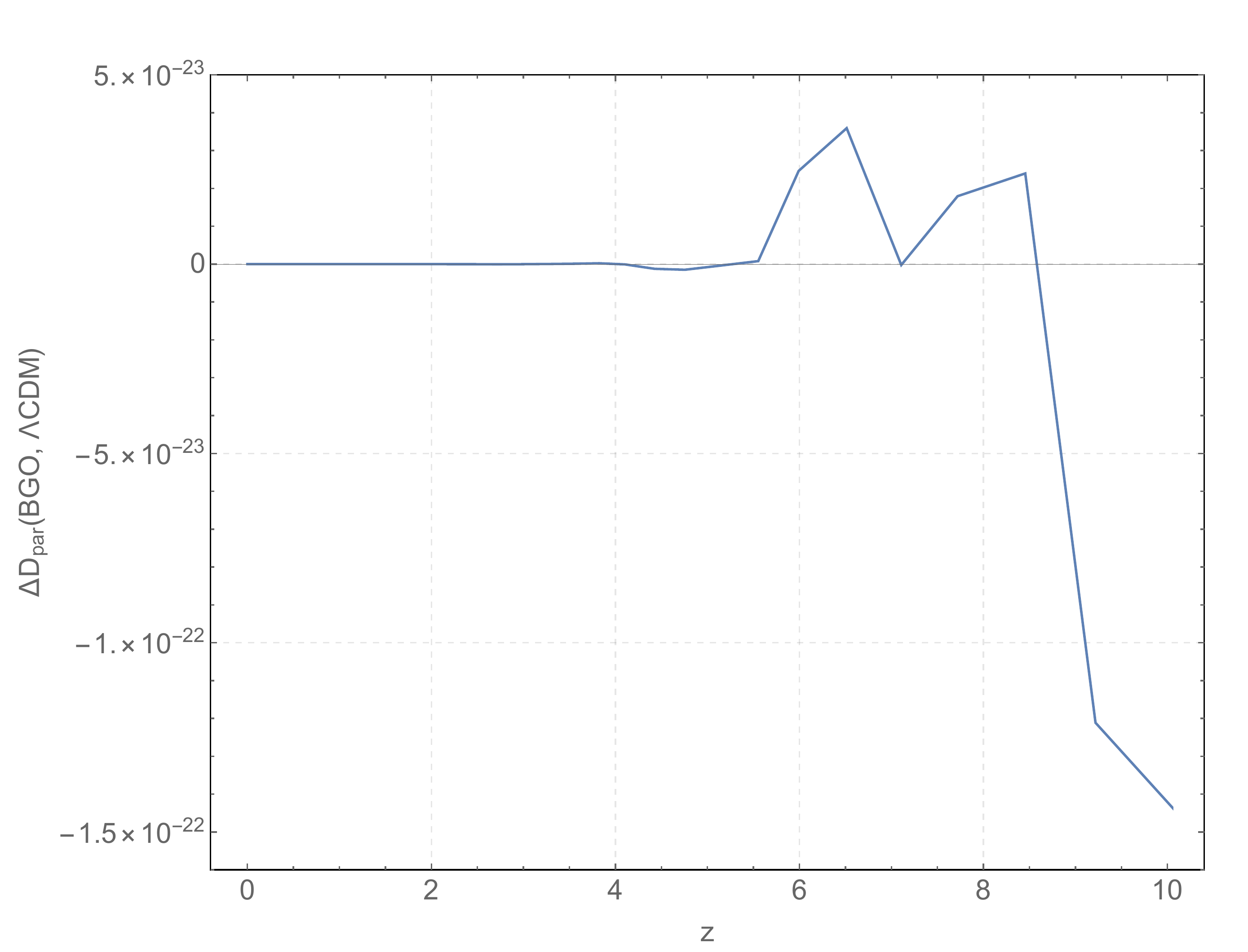}
        \caption{Parallax distance}
        \label{fig:LCDM_Dpar}
    \end{subfigure}
    \caption{Variations in the $\Lambda$CDM model, Eq.~\eqref{eq:deltaO_LCDM}, for the redshift~\ref{fig:LCDM_z}, the redshift drift~\ref{fig:LCDM_drift}, the angular diameter distance~\ref{fig:LCDM_Dang}, and the parallax distance~\ref{fig:LCDM_Dpar}. The variable in the horizontal axis is the redshift in $\Lambda$CDM. The values for the cosmological parameters are taken from \cite{planck2018param}.}\label{fig:LCDM}
\end{figure}

The fact that we have analytical expressions for the observables makes this model a perfect test-bed for the code.
We compare the results from {\tt BiGONLight} with the one in Eq.~\eqref{eq:z_FLRW}-\eqref{eq:z_drift_FLRW} by considering the variation
\begin{equation}
\Delta O({\rm BGO, \Lambda CDM}) \equiv \dfrac{O^{\rm BGO}-O^{\rm  \Lambda CDM}}{O^{\rm  \Lambda CDM}}\, .
\label{eq:deltaO_LCDM}
\end{equation}
As shown in Fig.~\ref{fig:LCDM}, the numerical implementation of the BGO method is in excellent agreement with the analytical formulas in $\Lambda$CDM, the variation $\Delta O$ being $10^{-22}\, \div \,10^{-32}$. The maximum value of $\Delta O$, of the order of $10^{-22}$, represents the numerical error over the observables and we reached such small values by using the precision control options {\tt WorkingPrecision, PrecisionGoal} and {\tt AccuracyGoal} implemented in {\tt Mathematica}.


\subsection{The Szekeres model}
\label{sec:Szekeres}
The second group of code tests is performed considering an inhomogeneous cosmological model which is an exact solution of Einstein equations, firstly presented in \cite{szekeres1975class}. The line element for the Szekeres spacetime is:
\begin{equation}
ds^2=- dt^2+e^{2 \alpha} {dx^{\rm 1}}^2+e^{2 \beta} ({dx^{\rm 2}}^2+{dx^{\rm 3}}^2)
\label{eq:Sz_generic}
\end{equation}
with $\alpha$ and $\beta$ functions of the spacetime coordinates $(t, x^{\rm 1},x^{\rm 2},x^{\rm 3})$ that are determined by solving the Einstein equations. We can distinguish two different families of Szekeres models depending whether $\partial_{\rm x^{\rm 3}} \beta \neq 0$ or $\partial_{\rm x^{\rm 3}} \beta = 0$: the first case defines the ``class I'' family, which is a generalization of LTB models (with non-concentric shells), while the case $\partial_{\rm x^{\rm 3}} \beta = 0$ corresponds to a simultaneous generalization of the Friedmann and Kantowski-Sachs models and it defines the ``class II'' family. For a cosmological formulation of the Szekeres spacetimes see e.g. \cite{Goode:1982pg}.
For our tests, we will use a class II Szekeres model filled with dust and a positive cosmological constant as presented in~\cite{Meures:Szekeres}. For this model, the line element \eqref{eq:Sz_generic} is rewritten as:
\begin{equation}
ds^2=a(\eta)^2 \left[- d\eta^2 + {dx^{\rm 1}}^2+{dx^{\rm 2}}^2 +Z(\eta,x^{\rm 1}, x^{\rm 2}, x^{\rm 3})^2 {dx^{\rm 3}}^2\right]
\label{eq:Sz_line}
\end{equation}
and, for the special case of axial symmetry around $x^{\rm 3}$, we have the following form for the function $Z$:
\begin{equation}
Z(\eta,x^{\rm 1},x^{\rm 2},x^{\rm 3})=1+\beta_{\rm +}(x^{\rm 3}) \mathcal{D}(\eta)+\beta_{\rm +}(x^{\rm 3}) B \left({x^{\rm 1}}^2+{x^{\rm 2}}^2\right)\, .
\label{eq_met_func}
\end{equation}
The function $\mathcal{D}$ is the growing mode solution of the first-order Newtonian density contrast defined as $\delta=\dfrac{\rho -\rho_{\rm \Lambda CDM}}{\rho_{\rm \Lambda CDM}}$ and it is given by, see e.g. \cite{villa2016relativistic},
\begin{equation}
\mathcal{D} (\eta) = \frac{a}{\frac{5}{2}\Omega_{\rm{m 0}} } \sqrt{1+\frac{\Omega_{\rm{\Lambda}}}{\Omega_{\rm{m 0}}}a^3}\,  {}_2 F_{1} \left( \frac{3}{2}, \frac{5}{6}, \frac{11}{6}, -\frac{\Omega_{\rm{\Lambda 0}}}{\Omega_{\rm{m 0}}}a^3 \right)\,,
\label{eq:grow_mode}
\end{equation}
with ${}_2 F_{1} \left(a,b,c, y\right)$ being the Gaussian (or ordinary) hypergeometric function. The term $B$ in Eq.~\eqref{eq_met_func}
 is constant and given by (see App. C in \cite{Grasso:2021zra})
\begin{equation}
B= \frac{5}{4} \stuff \frac{\cal D_{\rm in}}{a_{\rm in}}\, ,
\label{eq:link_on_B}
\end{equation}
where $\mathcal{D}_{\rm in}= a_{\rm in}$ for initial conditions set deeply in the matter-dominated era. The function $\beta_{\rm +}$ specifies the spatial distribution of the first-order density contrast and it can be related to the peculiar gravitational potential $\phi_{\rm 0}$ via the cosmological Poisson equation at present time
\begin{equation}
\beta_{\rm +}=-\dfrac{2}{3}\dfrac{\partial^2_{\rm x^{\rm 3}}\phi_{\rm 0}}{\stuff}\, .
\end{equation}
For the tests, we will use a sinusoidal profile for the peculiar gravitational potential $\phi_{\rm 0}= \mathcal{A} \sin(\omega x^{\rm 3})$ with $\omega=\dfrac{2 \pi}{500\, \rm Mpc}$ and amplitude $\mathcal{A}$ such that $\delta_{\rm 0}=0.1$ for the density contrast today. 

In the following, we will present the tests over the redshift, the angular diameter distance and the redshift drift. Contrary to the $\Lambda$CDM case, here the observables are obtained using two numerical methods and the difference is expressed by
\begin{equation}
\Delta O({\rm BGO, Sz}) \equiv \dfrac{O^{\rm BGO}-O^{\rm  Sz}}{O^{\rm  Sz}}\, .
\label{eq:deltaO_Sz}
\end{equation}
All the three tests are done considering that the observer $\mathcal{O}$ is located at the origin of the reference frame, with coordinates $(\eta_0, 0,0,0)$, and she/he sees the light coming from a comoving distant source $\mathcal{S}$, with coordinates $(\eta, 0,0, x^3)$. The light beam is propagating along the $x^3$ axis, with tangent vector $\ell^{\mu}=(\ell^{0},0,0,\ell^{3})$.

The first observable we test is the redshift: for a photon travelling along the $x^{\rm 3}$-axis it is
\begin{equation}
1+z^{\rm Sz}=\dfrac{\left(a \ell^0\right)|_{\rm \mathcal{S}} }{ \left(a \ell^0\right)|_{\rm \calO}}\, ,
\label{eq:Szekeres_z}
\end{equation}
where $\ell^0$ is obtained by solving the geodesic equation
\begin{equation}
\dfrac{d \ell^0}{d \eta} = - \ell^0 \left(2 \mathcal{H} + \dfrac{\dot{Z}}{Z}\right)\, . \label{eq:l0_Szekeres}
\end{equation}
In the above expressions $Z$ is given in Eq.~\eqref{eq_met_func}, $a$ is the scale factor \eqref{eq:a_LCDM}, and $\mathcal{H}$ the Hubble parameter \eqref{eq:E(z)} of the $\Lambda CDM$ background.
The variation $\Delta z(\rm BGO, Sz)$ refers to the numerical solution of Eq.~\eqref{eq:l0_Szekeres} as opposite to the $3+1$ geodesic solved by {\tt BiGONLight}.

The second observable under analysis is the angular diameter distance $D_{\rm ang}$. The standard procedure to compute $D_{\rm ang}$ is solving the Sachs focusing equation 
\begin{equation}
\dfrac{d^2 D_{\rm ang}}{d \lambda^2}= -\left(|\sigma|^2 + \dfrac{1}{2 }R_{\mu \nu}\ell^{\mu}\ell^{\nu}\right)D_{\rm ang}\, ,
\end{equation}
where the initial conditions given at $\cal O$ (i.e. the focusing point) are $D_{\rm ang}|_{\cal O}=0 $ and $\left. \dfrac{d D_{\rm ang}}{d \lambda}\right|_{\cal O}=(\ell_{\sigma}u^{\sigma})_{\rm \calO}$, and $|\sigma|$ is the shear that in the Szekeres model Eqs.~\eqref{eq:Sz_line}-\eqref{eq_met_func} simply vanishes, as shown in~\cite{Meures:2011ke}.
Using the Einstein equations to express $R_{\mu \nu}\ell^{\mu}\ell^{\nu}$ in terms of the density contrast and using conformal time instead of the affine parameter, the focusing equation assumes the form
\begin{equation}
\ddot{D}^{\rm Sz}_{\rm ang}+\dfrac{\dot{\ell}^0}{\ell^0} \dot{D}^{\rm Sz}_{\rm ang}=-  \dfrac{3}{2}\dfrac{\mathcal{H}_{\rm 0}^2\Omega_{\rm m_0}}{a}(\delta+1) D_{\rm ang}^{\rm Sz}\, ,
\label{eq:focusing_eq}
\end{equation}
where the dots refers to derivatives respect to conformal time $\eta$ and the initial conditions at the observation point are $ \ D_{\rm ang} \left|_{\rm \calO}\right.=0$ and $ \dot{D}_{\rm ang} \left|_{\rm \calO}\right.=\dfrac{(\ell_{\sigma}u^{\sigma})_{\rm \calO}}{\ell^0_{\rm \calO}}$.
In synchronous-comoving gauge the density contrast along the geodesic comes directly from the continuity equation and reads
\begin{equation}
\delta= - \dfrac{\mathcal{D} \beta_{\rm +}}{Z}=\dfrac{\dfrac{2}{3}\dfrac{\partial^2_{\rm x^{\rm 3}}\phi_{\rm 0}}{\stuff}\mathcal{D}}{1-\dfrac{2}{3}\dfrac{\partial^2_{\rm x^{\rm 3}}\phi_{\rm 0}}{\stuff}\mathcal{D}-\dfrac{2}{3}\dfrac{\partial^2_{\rm x^{\rm 3}}\phi_{\rm 0}}{\stuff} B ({x^{\rm 1}}^2+{x^{\rm 2}}^2)}\, .
\label{eq:Sz_delta}
\end{equation}
In our estimator Eq.~\eqref{eq:deltaO_Sz} the angular diameter distance $D_{\rm ang}^{\rm Sz}$ is obtained by integrating  Eq.~\eqref{eq:focusing_eq}, whereas $D_{\rm ang}^{\rm BGO}$ is obtained from Eq.~\eqref{eq:D_ang_BGO} with {\tt BiGONLight}.

The last test concerns the calculation of the redshift drift, namely the secular variation of the redshift of the source. It was calculated for some inhomogeneous cosmological models, see e.g. \cite{quartin2010distinguishing, yoo2011redshift, Mishra2012,balcerzak2013redshift, mishra2014redshift}, but to our knowledge there is no expression for the Szekeres model considered here, thus we give in the following a short derivation.
Let us consider that during the proper time lapse $\delta \tau_{\mathcal{O}}$ the spacetime coordinates of the observer change from $x^{\mu}_{\mathcal{O}}=(\eta_{\mathcal{O}}, 0,0,0)$ to $X^{\mu}_{\mathcal{O}}=(\Theta_{\mathcal{O}}, 0,0,0)$. Similarly, in the corresponding proper time lapse $\delta \tau_{\mathcal{S}}$, the spacetime coordinates of the source change from $x^{\mu}_{\mathcal{S}}=(\eta_{\mathcal{S}}, 0,0,x^3)$ to $X^{\mu}_{\mathcal{S}}=(\Theta_{\mathcal{S}},0,0,x^3)$. 
Note that in this gauge $\mathcal{O}$ and $\mathcal{S}$ are comoving, meaning that they both have fixed spatial positions (i.e. $\delta x^i_{\mathcal{O}} = \delta x^i_{\mathcal{S}}=0$), but the time changes differently at $\mathcal{O}$ and at $\mathcal{S}$ (i.e. $\Theta_\mathcal{O} - \eta_\mathcal{O} \neq \Theta_\mathcal{S} - \eta_\mathcal{S}$). The redshift and the conformal time of the source change according to\footnote{On the R.H.S. of Eqs.~\eqref{eq:Sz_Z(t)}-\eqref{eq:Sz_T(t)} we use the fact that the source location $x^i_{\mathcal{S}}$ has components on the $x^3$ axis only.}
\begin{align}
\mathcal{Z}(\Theta_{\mathcal{S}}, x^{i}_{\mathcal{S}})&=z(\eta_{\mathcal{S}} ,x^3_{\mathcal{S}})+\delta z(\eta_{\mathcal{S}} ,x^3_{\mathcal{S}})\label{eq:Sz_Z(t)}\\
\Theta(x^{i}_{\mathcal{S}})&=\eta(x^3_{\mathcal{S}})+\delta \eta(x^{3}_{\mathcal{S}})\, ,\label{eq:Sz_T(t)}
\end{align}
while the same quantities at the observer position $X^{\mu}_{\mathcal{O}}$ are $\mathcal{Z}(X^{\mu}_{\mathcal{O}})=z(x^{\mu}_{\mathcal{O}})=0 $ and $ \Theta(X^{i}_{\mathcal{O}})=\eta(x^{i}_{\mathcal{O}})+\delta \eta(x^{i}_{\mathcal{O}})$.
Our final aim is to compute the redshift drift, namely 
\begin{equation}
\zeta=\dfrac{\delta \ln(1+z)}{\delta \tau_{\mathcal{O}}}\, .
\label{eq:Sz_def_drift}
\end{equation}
Let us begin with the variation of the source redshift with respect to the observer proper time  $\frac{\delta z}{\delta \tau_{\mathcal{O}}}$, where it is better to obtain first $\frac{d \delta z}{d x^3}$ (Eq.~\eqref{eq:Sz_delz_dx1}) and $\frac{d \delta \tau}{d x^3}$ (Eq.~\eqref{eq:Sz_deltau_dx}) separately, and then combine them to get an ODE (Eq.~\eqref{eq:z_drift_Szekeres}), whose solution gives the redshift drift.
The starting point are the differentials $\frac{d \eta}{d x^3}$ and $\frac{d z}{d x^3}$. The first is simply given by the null condition $\ell^0=-Z \ell^3$, and reads
\begin{equation}
 \frac{d \eta}{d x^3}=-Z\, ,
\label{eq:Sz_deta_dx}
\end{equation}
while $\frac{d z}{d x^3}$ is obtained by differentiating Eq.~\eqref{eq:Szekeres_z}\footnote{From now on we drop the index $\mathcal{S}$, since  all the quantities are evaluated at the source.}
\begin{align}
\dfrac{d z}{d x^3}=\dfrac{1}{ \left(a \ell^0\right)|_{\rm \calO}}\left(\frac{d a}{d x^3} \ell^0 +\frac{d \ell^0 }{d x^3} a \right)|_{\rm \calS}&=\dfrac{\left(a \ell^0\right)|_{\rm \calS}}{ \left(a \ell^0\right)|_{\rm \calO}}\left(\frac{1}{a}\frac{d a}{d \eta}\frac{d \eta}{d x^3} +\frac{1}{\ell^0}\frac{d \ell^0 }{d \eta}\frac{d \eta}{d x^3} \right)|_{\rm \calS} \nonumber \\
&=(1+z)\left(\mathcal{H} Z + \dot{Z} \right)\, ,
\label{qe:Sz_dz_dx} 
\end{align}
where we used Eq.~\eqref{eq:Sz_deta_dx}, Eq.~\eqref{eq:l0_Szekeres}, and the fact that the redshift at the observer is fixed.
Now we use Eq.~\eqref{eq:Sz_deta_dx} to differentiate Eq.~\eqref{eq:Sz_T(t)}
\begin{align}
\frac{d \delta \eta}{d x^3}=\frac{d \Theta}{d x^3}-\frac{d \eta}{d x^3}=-Z(\Theta, x^3)+Z(\eta, x^3)&=-\left[Z(\eta, x^3)+\dot{Z}(\eta, x^3) \delta \eta\right]+Z(\eta, x^3) \nonumber \\
&=-\dot{Z} \delta \eta\, ,
\label{eq:Sz_deleta_dx}
\end{align} 
and similarly for the redshift we have
\begin{equation}
\dfrac{d \delta z}{d x^3}=\dfrac{d \mathcal{Z}}{d x^3}-\dfrac{d z}{d x^3}=(1+z)\left( Z \mathcal{H}+\dot{Z}\right)^{\cdot}\delta \eta +\left( Z \mathcal{H}+\dot{Z}\right)\delta z \, ,
\label{eq:Sz_delz_dx1}
\end{equation} 
where we keep first-order terms in $\delta \eta$ and $\delta z$ only.
Rearranging terms and using Eq.~\eqref{qe:Sz_dz_dx} again, we get
\begin{equation}
\dfrac{d }{d x^3}\left(\dfrac{\delta z}{1+z} \right)=\left( Z \mathcal{H}+\dot{Z}\right)^{\cdot}\delta \eta\, . \label{eq:Sz_delz_dx2}
\end{equation}
The final step is to express the variation of the conformal time $\delta \eta$ in terms of the proper time at the observer $\delta \tau_{\mathcal{O}}$. We use their relation, which is simply
\begin{equation}
\delta \tau=\sqrt{|g_{\mu \nu} \delta x^{\mu} \delta x^{\nu}|}=a \sqrt{|-\delta \eta^2 + Z^2 \delta {x^{\rm 3}}^2|}= a \delta \eta\, ,
\label{eq:Sz_tau_eta}
\end{equation}
since $\delta x^i=0$. We need the derivative with respect to $x^3$, which reads
\begin{equation}
\frac{d \delta \tau}{d x^3}=\frac{d (a \delta \eta)}{d x^3}=\frac{d a}{d x^3}\delta \eta +a \frac{d \delta \eta}{d x^3}=\left( \frac{1}{a} \frac{d a}{d \eta}\frac{d \eta }{d x^3} - \dot{Z}\right) a \delta \eta=-\left( \mathcal{H} Z + \dot{Z}\right) \delta \tau\, ,
\label{eq:Sz_deltau_dx}
\end{equation}
and, together with Eq.~\eqref{qe:Sz_dz_dx} and Eq.~\eqref{eq:Sz_tau_eta},   the solution of the last equality is
\begin{equation}
\delta \eta=\frac{1}{a}\dfrac{\delta \tau_{\mathcal{O}}}{1+z}\, . \label{Sz_eta}
\end{equation}
By changing $x^3$ to conformal time $\eta$ with the chain rule $\frac{d}{dx^3}=\frac{d\eta}{dx^3}\frac{d}{d\eta}=-Z\frac{d}{d\eta}$, we get from Eq.~\eqref{eq:Sz_delz_dx2} the ODE in $\eta$ for the redshift drift in the Szekeres model
\begin{equation}
\frac{d\zeta}{d \eta}\equiv\frac{d}{d \eta}\left(\frac{\delta \log(1+z)}{\delta \tau_{\mathcal{O}}}\right)=-\frac{1}{a(1+z)}\frac{\left(H Z+ \dot{Z}\right)^{\cdot}}{Z}\, .
\label{eq:z_drift_Szekeres}
\end{equation}
For our test, in the estimator Eq.~\eqref{eq:deltaO_Sz}, $\zeta^{\rm Sz}$ is the numerical solution of Eq.~\eqref{eq:z_drift_Szekeres} and $\zeta^{\rm BGO}$ is the expression in Eq.~\eqref{eq:z_DRIFT_BGO} obtained with {\tt BiGONLight}.
\begin{figure}[ht]
    \centering
    \begin{subfigure}{0.49\linewidth}
        \includegraphics[width=\linewidth]{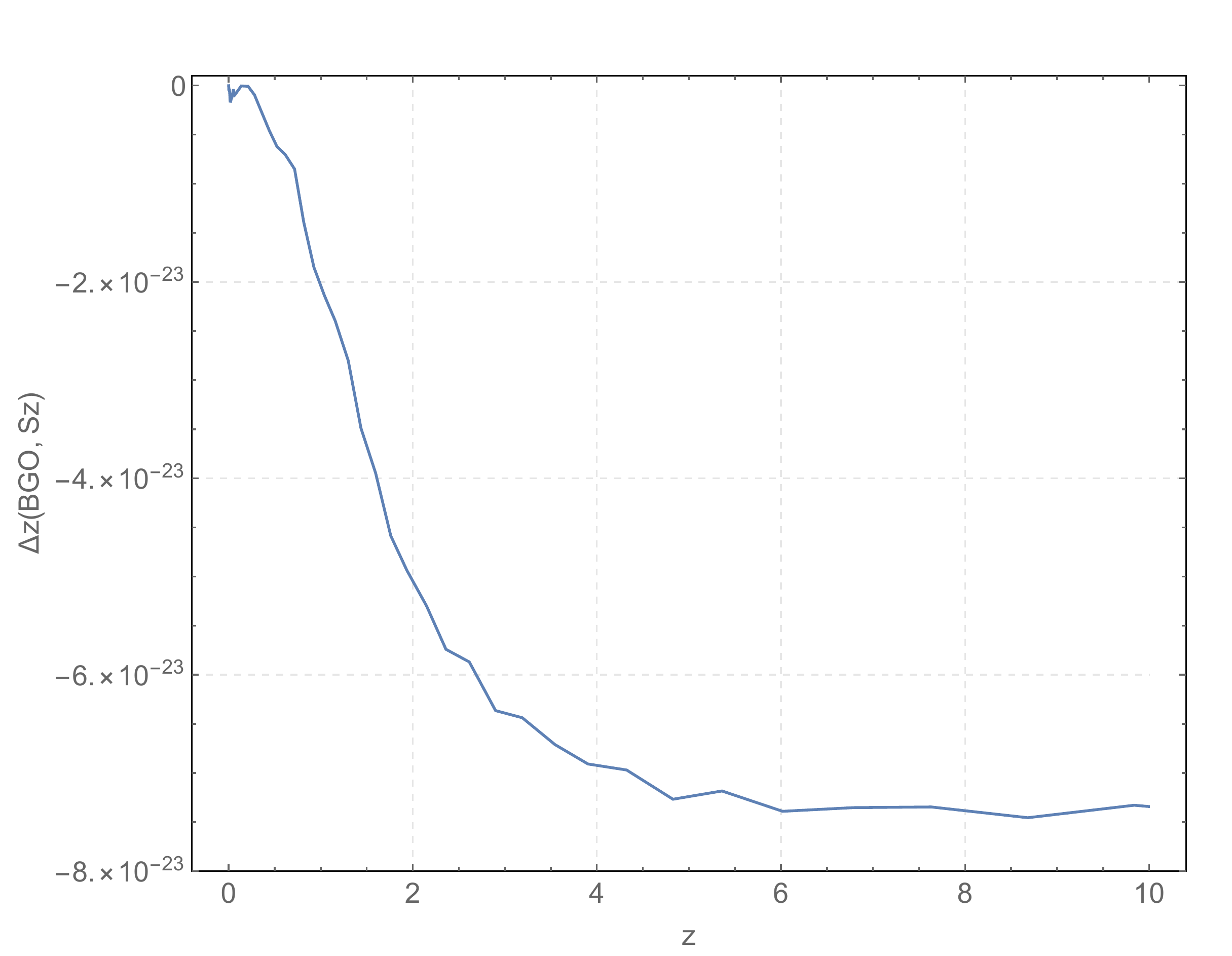}
       \caption{Redshift}
        \label{fig:Sz_z}
    \end{subfigure}
    \begin{subfigure}{0.49\linewidth}
        \includegraphics[width=\linewidth]{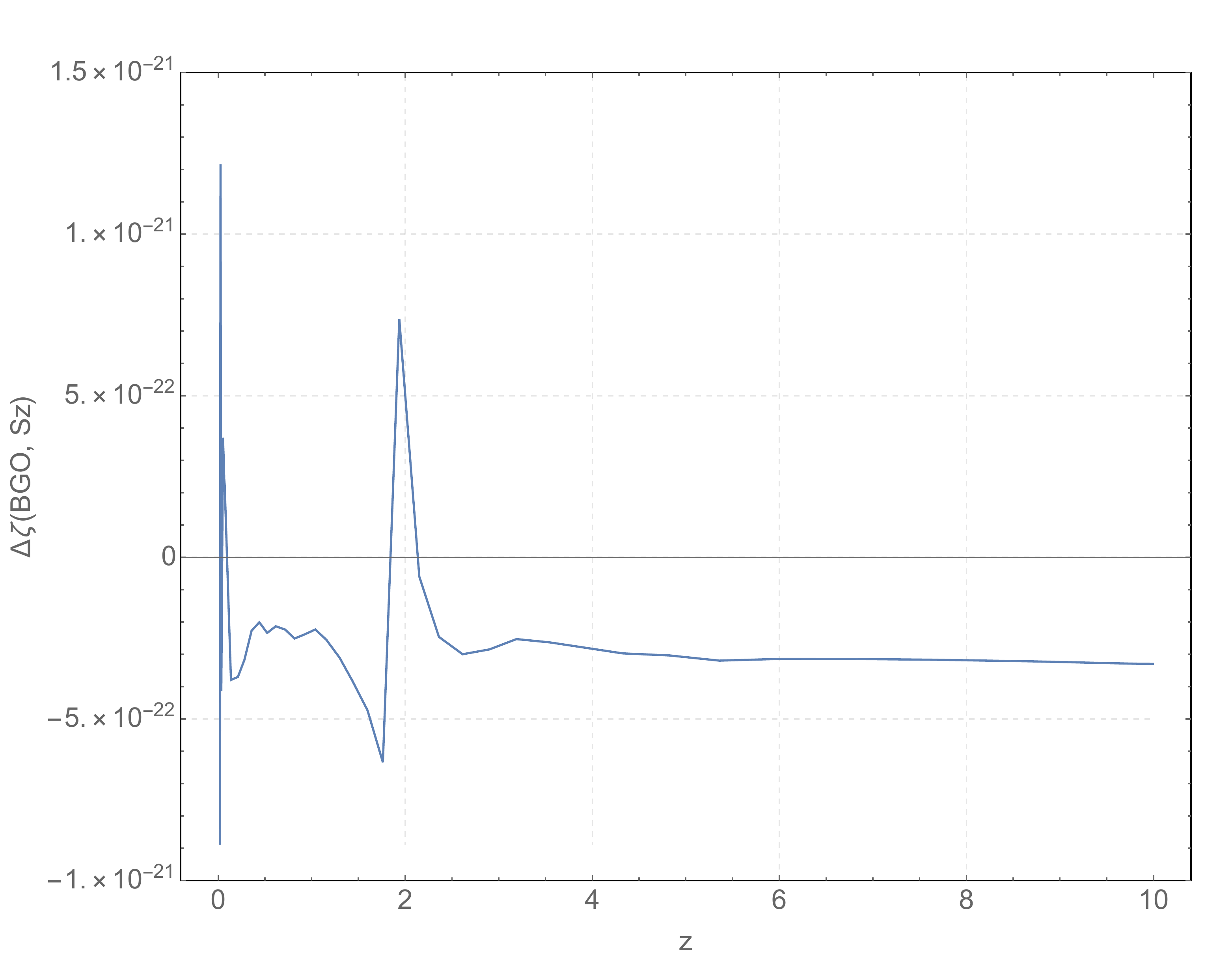}
        \caption{Redshift drift}
        \label{fig:Sz_drift}
    \end{subfigure}
    \begin{subfigure}{0.49\linewidth}
        \includegraphics[width=\linewidth]{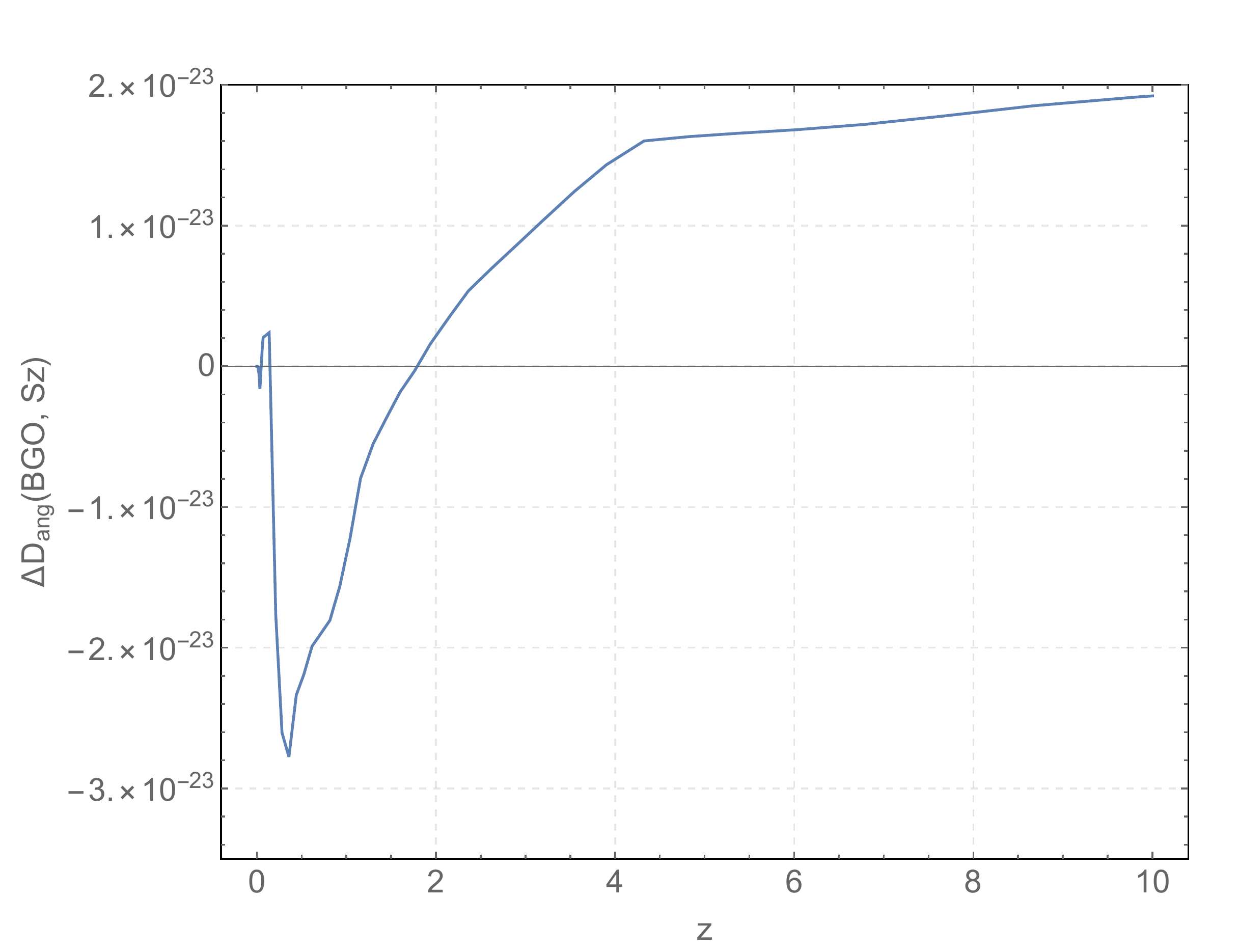}
        \caption{Angular diameter distance}
        \label{fig:Sz_Dang}
    \end{subfigure}
    \caption{Variations in the Szekeres model, Eq.~\eqref{eq:deltaO_Sz}, for the redshift~\ref{fig:Sz_z}, the redshift drift~\ref{fig:Sz_drift}, and the angular diameter distance~\ref{fig:Sz_Dang}. The variable in the horizontal axis is the redshift in the Szekeres model.}\label{fig:Sz}
\end{figure}


As for the $\Lambda$CDM model, also for the Szekeres model we have a very good agreement between the observables calculated using  the standard procedure and the observables from \texttt{BiGONLight}. The smallness of all the variations $\Delta z (\rm BGO, Sz)$, $\Delta \zeta(\rm BGO, Sz)$ and $\Delta D_{\rm ang}(\rm BGO, Sz)$ shown in Fig.~\ref{fig:Sz} means that our code could be a reliable tool for light propagation in inhomogeneous cosmologies, represented here with the Szekeres model, which is computationally more complicated than the homogeneous $\Lambda$CDM case.


\subsection{A dust universe in Numerical Relativity}
\label{sec:ET}
The main application of the {\tt BiGONLight} package is the computation of observables from numerically simulated spacetimes. 
 For our test we choose to use the {\tt FLRWSolver}\footnote{{\color{blue}{\tt {https://github.com/hayleyjm/FLRWSolver\_public}}.}}, \cite{macpherson2017}, which is a module (or {\it thorn}) of the {\tt Einstein Toolkit} (ET), \cite{loffler2012einstein}: the ET is a collection of  open-source codes, called thorns, based on the {\tt Cactus} framework, \cite{goodale2002cactus}, which allows to solve the Einstein equations in the BSSN formulation of the 3+1 splitting, \cite{shibata1995evolution, baumgarte1998numerical}. The role of the {\tt FLRWSolver} is to provide the initial conditions in the form of linear perturbations around the Einstein-de Sitter (EdS) background, which are then evolved with the ET. Here, we limit ourself to the EdS background model and set perturbations to zero. In other words, we consider a FLRW model in which the Universe is flat and contains only cold dark matter.  The line element of the EdS model in conformal time is
\begin{equation}
ds^2={a^2_{\rm EdS}}(\eta) \left(-d\eta^2+ {dx^2}^2 + {dx^2}^2 + {dx^3}^2 \right)\, ,
\end{equation}
where $a_{\rm EdS}(\eta)=\eta^2$ is the scale factor. 
We carry out the simulation in a cubic domain $-L \le \{x^1, x^2, x^3 \}\le L$ with periodic boundary conditions and spatial resolution $\Delta x=\Delta y=\Delta z= \frac{L}{20}$, where $L$ is the simulation unit length in Mpc\footnote{The physical value is $L=268.11\, {\rm Mpc}$, as it is explained in \ref{apx:Units}.}. The initial data are given at $\eta_{\rm in}=L$ and such that $\gamma^{\rm in}_{i j}=\delta_{i j}$. 
 The simulation runs with the ET up to $\eta_0=33.2 L$, which corresponds to integrating from redshift $z = 1100$ to present time $z=0$, and we choose a fixed temporal resolution $\Delta \eta=\frac{L}{100}$ due to computational time convenience.
To give an estimation of the simulation error we define
\begin{equation}
\Delta a({\rm ET, EdS}) \equiv \dfrac{a^{\rm ET}-a^{\rm EdS}}{a^{\rm EdS}}\, ,
\label{eq:DeltaA}
\end{equation}
which is the variation between the analytical scale factor in EdS, i.e. $a^{\rm EdS}=\eta^2$ and the scale factor from the numerical simulation $a^{\rm ET}={\rm det}(\gamma_{i j})^{\frac{1}{6}}$.
\begin{figure}[ht]
    \centering
        \includegraphics[width=0.8\linewidth]{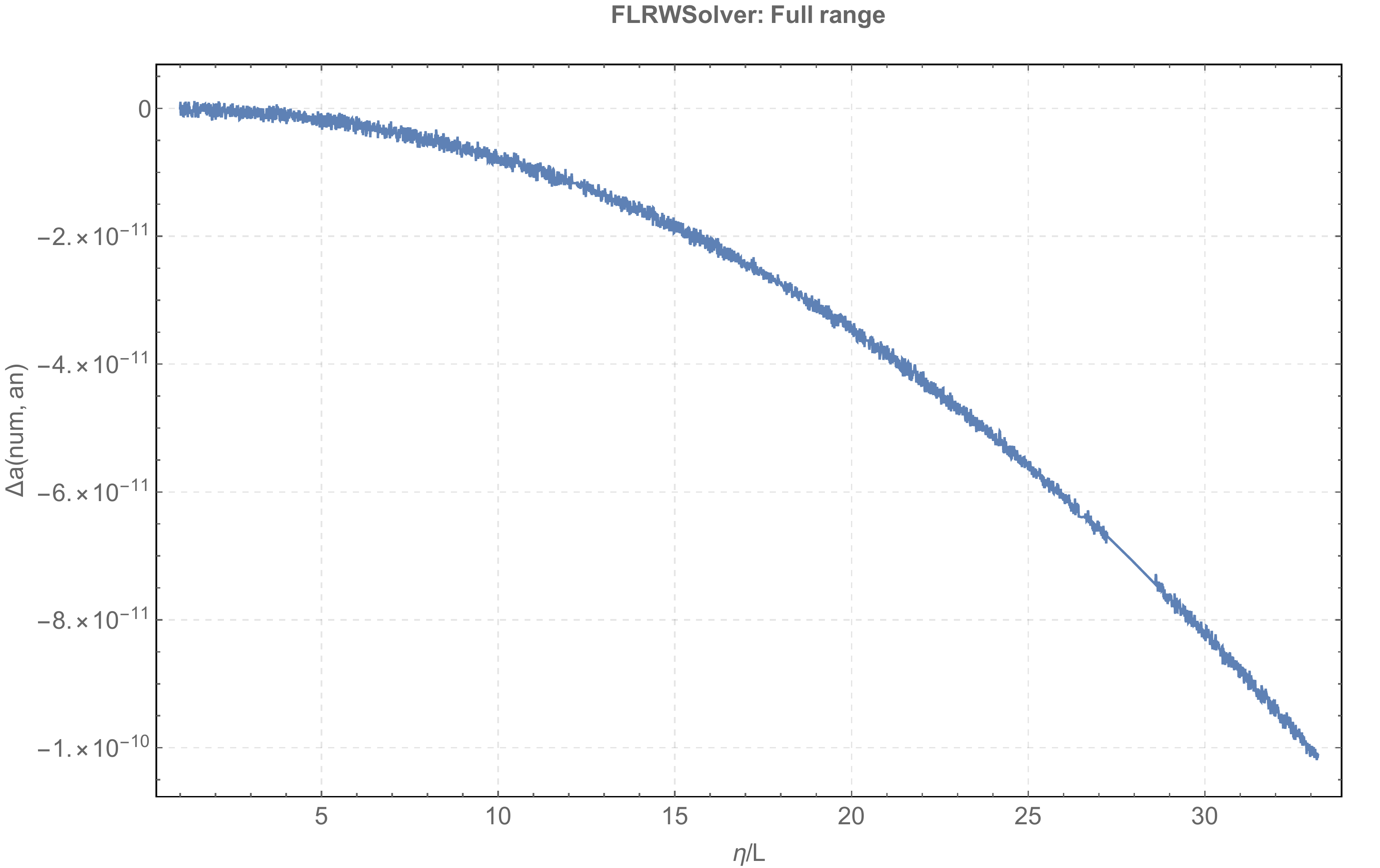}
\caption{Scale factor in EdS: ET simulation precision. The variable on the horizontal axis is the conformal time in computational units (see \ref{apx:Units}).}
    \label{fig:scale_ET}
\end{figure}
The result, shown in Fig.~\ref{fig:scale_ET}, is of the order of $ 10^{-10}$ and this value is determined by the specific setting that we choose for the ET simulation.

The simulated EdS model constitutes the playground of our tests. We perform light propagation with {\tt BiGONLight} using forward integration in time with the method described in Sec.~\ref{sec:W3+1}. We start from the source $\mathcal{S}$, placed at redshift $z= 10$ in $x^{\mu}_{\calS}=(\eta_{\calS}, 0, 0, 0)$, and we end at the observer $\calO$. The emitted light moves along the diagonal of the cubic domain with initial tangent vector $\ell^{\mu}_{\calS}=(-1, -\frac{1}{2}, \frac{1}{2},\frac{\sqrt{2}}{2})$, until it reaches the observer at $x^{\mu}_{\calO}$. 
The numerical accuracy in the calculation of the observables is tested by means of the variation $\Delta O({\rm BGO, ET})$ defined as
\begin{equation}
\Delta O({\rm BGO, EdS}) \equiv \dfrac{O^{\rm BGO}-O^{\rm  EdS}}{O^{\rm  EdS}}\, ,
\label{eq:deltaO_ET}
\end{equation}
where $O^{\rm BGO}$ is computed numerically with {\tt BiGONLight} using as input the EdS model simulated with the ET and $O^{\rm EdS}$ is the analytical expression in the EdS model that reads
\begin{align}
D_{\rm ang}^{\rm EdS}&=\dfrac{2 a_0}{\mathcal{H}_{\rm 0}}\dfrac{\sqrt{1+z}-1}{(1+z)^{\frac{3}{2}}} \label{eq:ET_Dang}\, , \\
D_{\rm par}^{\rm EdS}&=\dfrac{a_0}{\mathcal{H}_{\rm 0}} \dfrac{\sqrt{1+z}-1}{\frac{3}{2}\sqrt{1+z}-1} \label{eq:ET_Dpar}\, , \\
\zeta^{\rm EdS}&=\dfrac{\mathcal{H}_{0}}{a_0} (1-\sqrt{1+z}) \label{eq:ET_drift} \, .
\end{align}
These are obtained by integrating Eqs.~\eqref{eq:D_ang_FLRW}-\eqref{eq:z_drift_FLRW} with $\Omega_{\rm m_0}=1$ and $\Omega_{\rm \Lambda}=0$.
\begin{figure}[ht]
    \centering
    \begin{subfigure}{0.49\linewidth}
        \includegraphics[width=\linewidth]{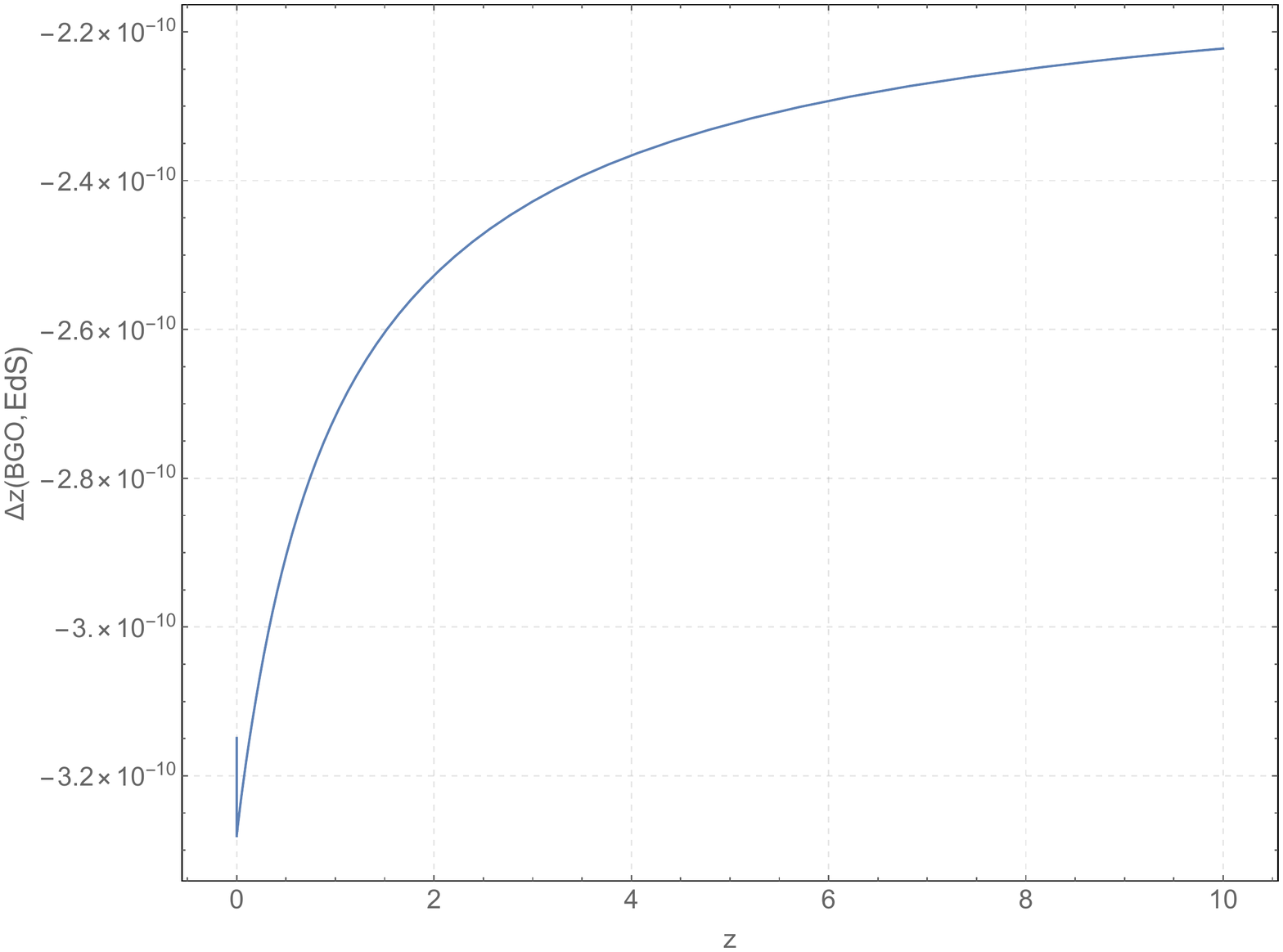}
       \caption{Redshift}
        \label{fig:ET_z}
    \end{subfigure}
    \begin{subfigure}{0.49\linewidth}
        \includegraphics[width=\linewidth]{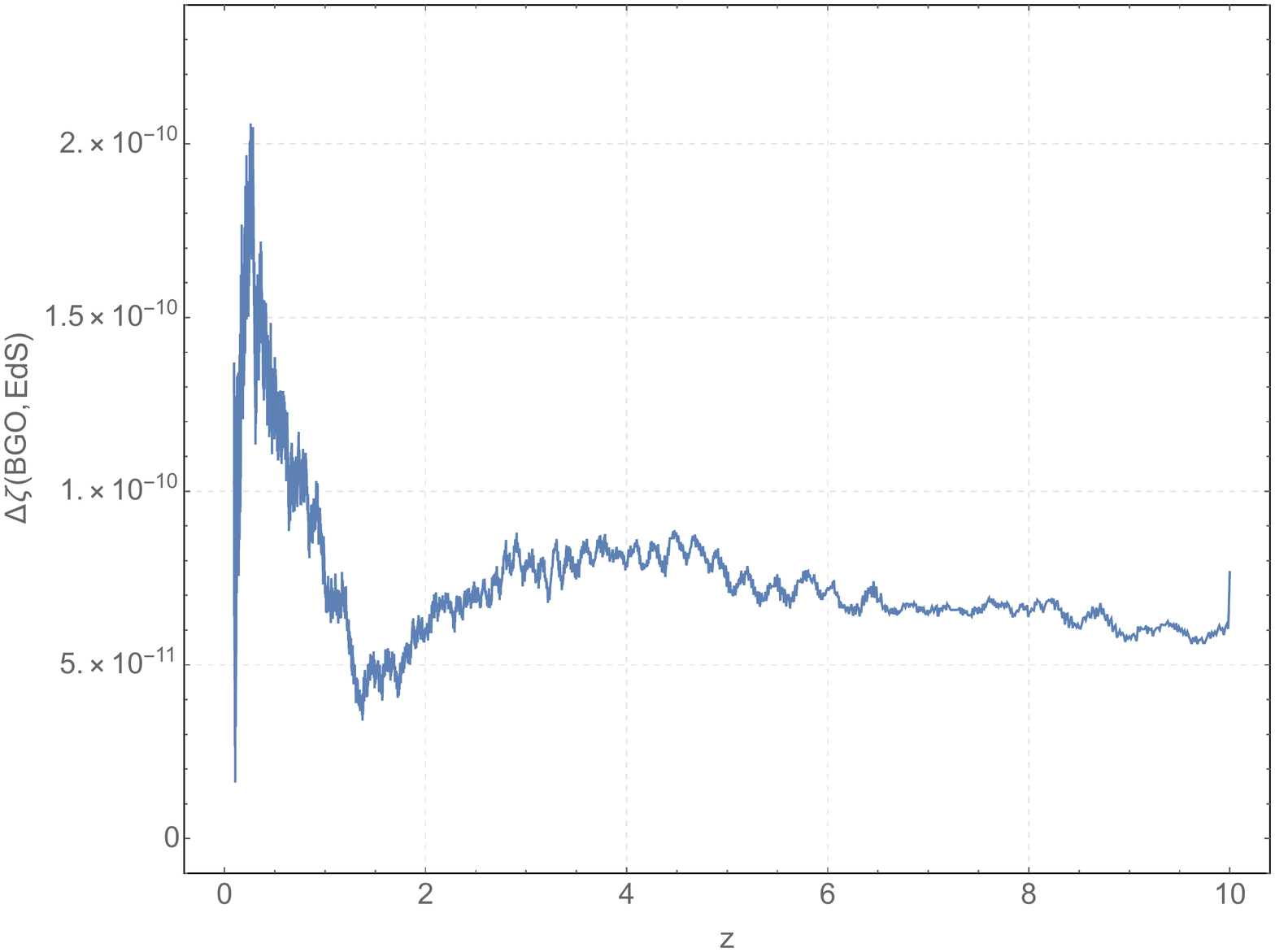}
        \caption{Redshift drift}
        \label{fig:ET_drift}
    \end{subfigure}
    \begin{subfigure}{0.49\linewidth}
        \includegraphics[width=\linewidth]{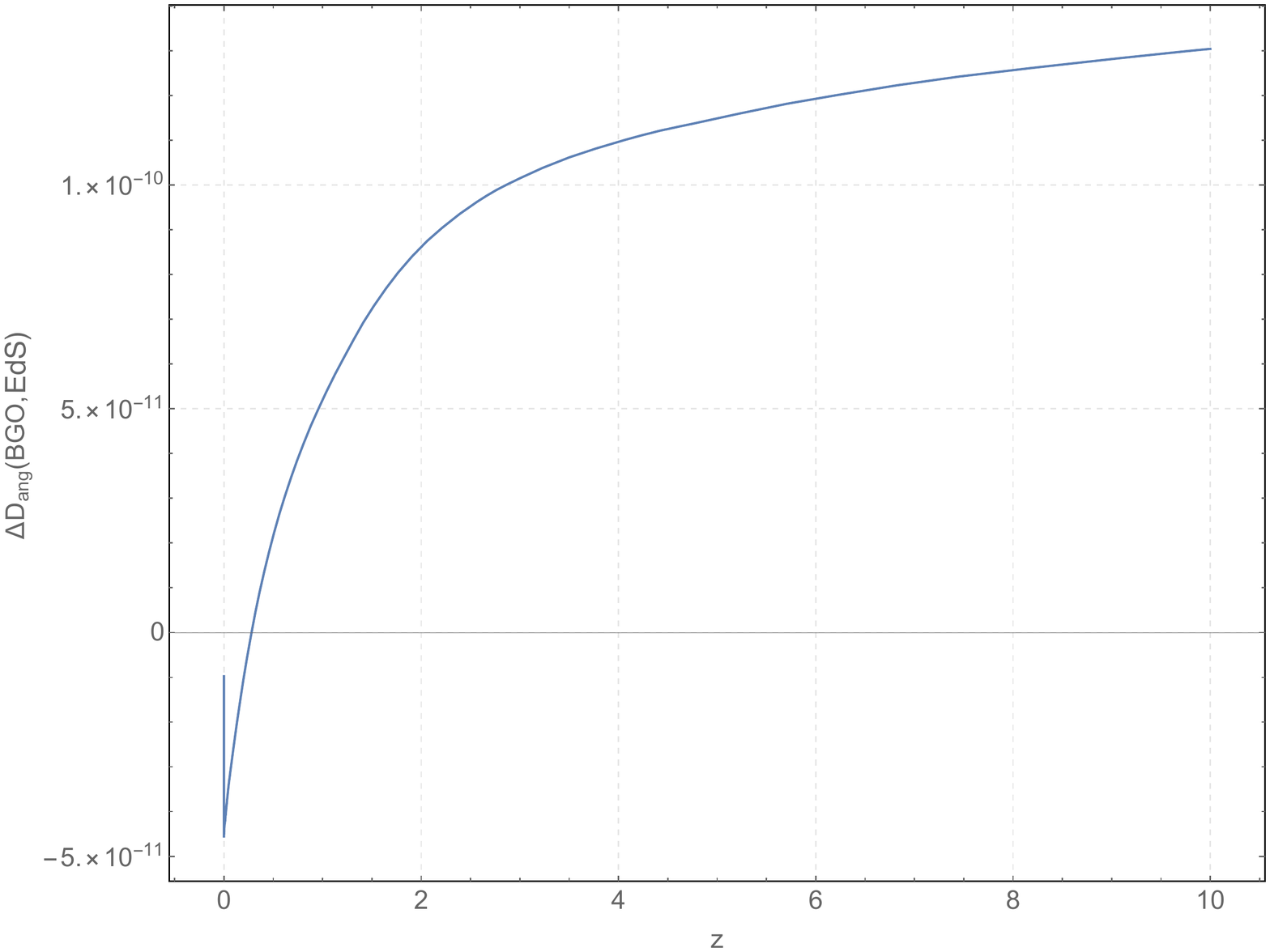}
        \caption{Angular diameter distance}
        \label{fig:ET_Dang}
    \end{subfigure}
    \begin{subfigure}{0.49\linewidth}
        \includegraphics[width=\linewidth]{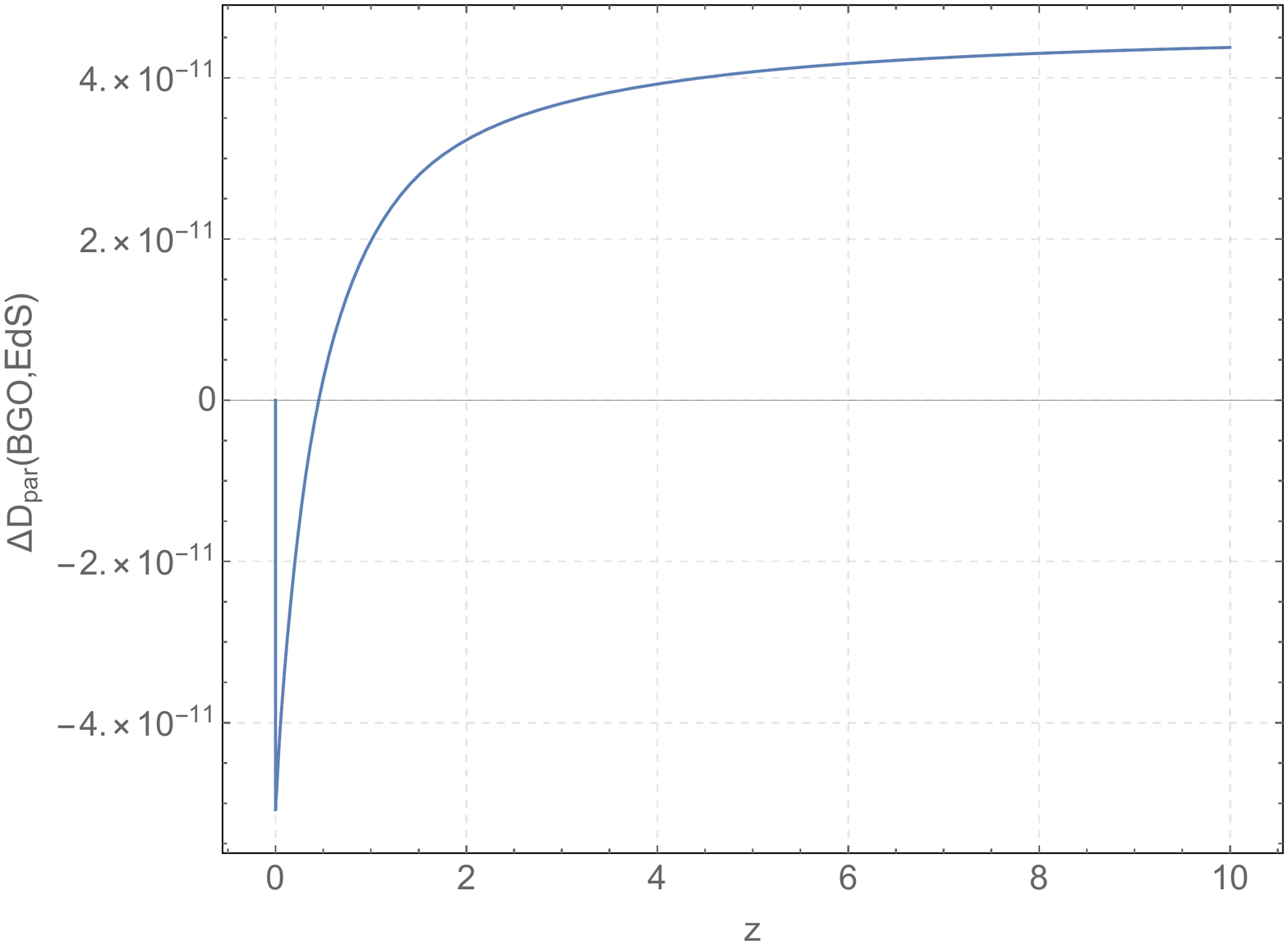}
        \caption{Parallax distance}
        \label{fig:ET_Dpar}
    \end{subfigure}
    \caption{Variations in the EdS model with ET, Eq.~\eqref{eq:deltaO_ET}, for the redshift~\ref{fig:ET_z}, the redshift drift~\ref{fig:ET_drift}, the angular diameter distance~\ref{fig:ET_Dang}, and the parallax distance~\ref{fig:ET_Dpar}. The variable in the horizontal axis is the redshift in EdS.}
    \label{fig:ET}
\end{figure}
The results are shown in Fig.~\ref{fig:ET}. 
What we see is the error on $\Delta O({\tt BGO, EdS})$ which has in principle two contributions: one from the simulation of the EdS model and the other from the simulation of light propagation. We have already isolated the second contribution coming from {\tt BiGONLight} in the $\Lambda$CDM test. Indeed, since we use the analytical solution for $\Lambda$CDM both in the numerical and analytical computation for the observables, the error $\Delta O({\tt BGO, \Lambda CDM})$ we find in Fig.~\ref{fig:LCDM} is entirely due to the simulation of light propagation and is of the order of $10^{-22} \div 10^{-31}$.
On the other hand, in Fig.~\ref{fig:scale_ET} we see that the accuracy of the ET simulation we use is much larger, i.e. of the order of $10^{-10}$, which is of the same order of the one for $\Delta O({\tt BGO, EdS})$ in Fig.~\ref{fig:ET}. Therefore we can conclude that the final error on the observables we find in Fig.~\ref{fig:ET} is settled by the accuracy of the ET in simulating the EdS model. Let us finally remark that for this test we perform light propagation using forward integration, namely from the source $\calS$ to the observer $\calO$. We also repeated the computation of $\Delta O({\tt BGO, EdS})$ using backward integration, form $\calO$ to $\calS$, in {\tt BiGONLight} and we found the same results.
\section{Conclusions}
\label{sec:concl}

In this paper we present and test {\tt BiGONLight}, a {\tt Mathematica} package for relativistic light propagation in Numerical Relativity. Our new code implements the BGO formalism, a new approach to geometric optics in General Relativity, firstly introduced in Ref.~\cite{grasso2019BGO}, which we recasted here in the $3+1$ form to be compatible with the structure of relativistic numerical simulations. The generality of the BGO framework allows {\tt BiGONLight} to be suitable to study light propagation in different contexts: from small scales, e.g. inside our Galaxy, to the ultra-large scales of Cosmology, as long as light propagation can be treated within the assumptions of geometric optics. {\tt BiGONLight} is extremely flexible also from the technical point of view, requiring as input only the spacetime metric and the observer and source kinematics: the user can choose any gauge and also the type of input to use, namely numerical from a relativistic simulation or analytical from an exact solution of the Einstein field equations. An analytical solution for the spacetime metric in any perturbative scheme can be also given as input but the code does not apply the perturbative scheme to light propagation. This is indeed the approach we adopted in Ref.~\cite{Grasso:2021zra}. We plan to implement perturbation theory in a future version of {\tt BiGONLight}.
A key feature of the BGO framework is that all optical effects are encoded in the bi-ejective $\mathcal{W}$ map, which can be written form the observer $\calO$ to the source $\calS$ or viceversa. In this paper we give the ready-to-use transformation relations in Eqs.~\eqref{eq:WXX_inverse}-\eqref{eq:WLL_inverse}. This property in {\tt BiGONLight} means that the user is free to adopt different approaches to trace light propagation. On one hand, one can trace the photons from the observer to the source using the conventional backward integration, on the other hand one can choose to do the other way around and use forward integration. This second method is particularly convenient in Numerical Relativity because it allows us to perform light propagation on-the-fly with a simulation of spacetime dynamics, which employs integration forward in time by construction.

We decide to test {\tt BiGONLight} in the computation of four cosmological observables: the redshift, the angular diameter distance, the parallax distance, and the redshift drift, the last two being real-time observables that are new in the cosmological context. 
We perform three different kinds of tests. In the first we test the accuracy of the code against the analytical results in the $\Lambda$CDM model and we find an excellent agreement of the order of $10^{-23} \div 10^{-31}$, see Fig.~\ref{fig:LCDM}. We are able to reach this extremely good accuracy by using the precision control options and the many well-tested numerical methods to solve ODE implemented in {\tt Mathematica}.
For the second test we consider the Szekeres inhomogeneous cosmological model as presented in Refs.~\cite{Meures:2011ke, Grasso:2021zra} and we compare the BGO formalism with the standard procedures used for the angular diameter distance and the redshift drift. Let us point out here that the BGO formalism provides a unified way to compute multiple observables within the same framework, contrary to the standard approach. In particular, for the angular diameter distance we compare the result obtained from the BGO, Eq.~\eqref{eq:D_ang_BGO}, with the one from the Sachs equation, Eq.~\eqref{eq:focusing_eq}. For the redshift drift we compare the BGO general formula, Eq~\eqref{eq:z_DRIFT_BGO}, with the result of the standard calculation, Eq.~\eqref{eq:z_drift_Szekeres}, that we derive here for the specific set up for light propagation in the Szekeres model considered. Also in this case we find a very good agreement of the order of $10^{-22}$, see Fig.~\ref{fig:Sz}. 
For our final test we consider the EdS background simulated with the {\tt Einstein Toolkit} together with the {\tt FLRWSolver}. The numerical output of this simulation is used to compute the observables with the BGO in {\tt BiGONLight} which are then compared with the usual analytical formulas. Our findings are shown in Fig.~\ref{fig:ET} where we see that the accuracy on the observables calculated with {\tt BiGONLight} is ruled by the accuracy of the {\tt Einstein Toolkit} simulation, which is limited to the order of $ 10^{-10}$ for the specifications that we use in our paper. Contrary to the first two tests, here the BGO are calculated by integrating the GDE forward in time and then we use the BGO property in Eq.~\eqref{eq:W(p,o)_from_W(p,s)} to obtain the observables, see Sec.~\ref{sec:obs}.
The current version of {\tt BiGONLight} is designed to perform light propagation in post-processing and not on-the-fly, with the advantage of being able to accept input from different numerical codes for cosmological dynamics.


\vspace{0.5 cm}

{\bf Acknowledgments:}\\
This work was supported by the National Science Centre, Poland (NCN) via 
the SONATA BIS programme, grant No~2016/22/E/ST9/00578 for the project
\emph{``Local relativistic perturbative framework in hydrodynamics and 
general relativity and its application to cosmology''}. We thank Miko\l{}aj Korzy\'nski for carefully reading the draft, Hayley Macpherson for useful comments and tips about the {\tt FLRWSolver}, and Hayley Macpherson and Julian Adamek for insightful discussions during the workshop \textit{``GR simulations in cosmology''}, Queen Mary University of London, $7$-$8$ September $2020$.


\section*{Appendix}
\appendix


\section{Units}
\label{apx:Units}

Throughout this article all quantities are expressed in geometric units, i.e. units defined by the relation $G = c = 1$, thus such that masses, time and lengths have the same unit of measurement. For instance, we can fix a unit of length\footnote{The same can be done by fixing a unit of time $T$ to define $[mass]=T \frac{c^3}{G}$ as unit of mass and $[length]=T c$ as unit of length, or fixing a unit of mass $M$ to define $[time]=M \frac{G}{c^3}$ as unit of time and $[length]=M \frac{G}{c^2}$ as unit of length.} $[length]=L$ and define $[mass]=\frac{c^2 L}{G}$ as unit of mass and $[time]=\frac{L}{c}$ as unit of time: in geometric units they all reduce to $[mass]=[time]=[length]=L$. Within this choice, every physical quantity $Q_{\rm phys}$ can be expressed as $Q_{\rm phys}=Q_{\rm comp} L^{\alpha}$, where $Q_{\rm comp}$ is dimensionless, $L$ is a arbitrary length to be chosen, and $\alpha$ is a certain exponent. This way of writing is particularly useful in numerical simulations, where all physical quantities are represented as dimensionless numbers and units are assigned when analysing the results.
Usually $L$ is fixed to be a length meaningful for the specific physical situation under consideration. For example, a common choice in numerical cosmological dynamics is to set $L$ equal to a characteristic length of the simulation (e.g. N-body and GR hydrodynamics), such as the side of the simulated box. 
In cosmology it is usually chosen to set $L$ equal to the conformal time in Mpc, i.e. $L=\eta^{\rm phys}$, thus $\eta^{\rm comp}=1$, at some special instant like e.g. at initial time or today. This choice is particularly convenient, since  conformal time is found by integrating the Friedmann equation and reads
 \begin{equation}
 \eta =\dfrac{1}{H_0}  \mathlarger{\int}^{a}_{0} \dfrac{d \tilde{a}}{\tilde{a}^2E(\tilde{a})}\, ,
 \label{eq:UNITS_eta}
 \end{equation}
where $E(a)=\sqrt{\Omega_{\rm m_0}\left(\frac{a_0}{a}\right)^3+\Omega_{\rm \Lambda}}$ for a universe containing cold dark matter and a cosmological constant. 

For the $\Lambda$CDM model, Sec.~\ref{sec:LCDM}, and for the Szekeres model, Sec.~\ref{sec:Szekeres}, we choose $L=\eta_0$ and we normalize the scale factor to 1 today. This is the natural choice for this two cases since we studied light propagation backward in time.
By integrating Eq.~\eqref{eq:UNITS_eta} together with the normalization $a_0=1$ for the value of the today scale factor we have
\begin{equation}
\eta^{\rm \Lambda CDM} = \dfrac{1}{\mathcal{H}_0  3^{\frac{1}{4}} \Omega_{\Lambda}^{\frac{1}{6}} \Omega_{\rm m_0}^{\frac{1}{3}}} F\left( \arccos\left(\frac{1+(1-\sqrt{3})\sqrt[3]{\frac{\Omega_{\Lambda}}{\Omega_{\rm m_0}}}a}{1+(1+\sqrt{3})\sqrt[3]{\frac{\Omega_{\Lambda}}{\Omega_{\rm m_0}}}a}\right);\frac{\sqrt{2+\sqrt{3}}}{2}\right)\, ,
\label{eq:UNITS_eta_LCDM}
 \end{equation}
with $F(x;y)$ being the elliptic integral of the first kind. Substituting the values of the cosmological parameters from \cite{planck2018param}, $\Omega_{\rm m_0}=0.315$ and $\Omega_{\rm \Lambda}=0.685$ and expressing the Hubble constant in Mpc, i.e. $\mathcal{H_{\rm 0}}=2.2469 \times 10^{-4}\, {\rm Mpc^{-1}}$, Eq.~\eqref{eq:UNITS_eta_LCDM} gives $L=\eta^{\rm \Lambda CDM}_0=\eta^{\rm Sz}_0=14.4152\,  \text{Gpc}$.

For the EdS model, Sec.~\ref{sec:ET}, we study light propagation forward in time and we decided to use the same conventions as the one implemented in the ET. Here, the simulation is carried out in a cubic domain volume of comoving side $2L$, which is initialised at initial time  $\eta_{\rm in}$ (and not today) and the scale factor is normalised to 1 at $\eta_{\rm in}$. Let us start by integrating \eqref{eq:UNITS_eta} for the EdS model, i.e. $\Omega_{\rm m_0}=1$ and $\Omega_{\rm \Lambda}=0$, which gives
\begin{equation}
\eta^{\rm EdS}=\dfrac{2}{\mathcal{H}_0}\sqrt{\dfrac{a}{a_0}}\, .
\label{eq:UNITS_eta_EdS}
\end{equation}
The value of $L$ is set by evaluating Eq.~\eqref{eq:UNITS_eta_EdS} at initial time and choosing $\eta^{\rm EdS}_{\rm in}=L$. Substituting $a_{\rm in}=1$ and $\sqrt{a_0}=33.2$ from the numerical simulation, and $\mathcal{H}_{0}=2.2469 \times 10^{-4}\, {\rm Mpc^{-1}}$ from \cite{planck2018param} in Eq.~\eqref{eq:UNITS_eta_EdS}, it follows that the value of $L$ is
\begin{equation}
L=268.11 \, {\rm Mpc}\, .
\end{equation}

We report for completeness the dimensions of all the main quantities in units of the characteristic length $L$
\begin{table}[ht!]
\centering
\begin{tabular}{|l|l|}
\hline
 \multicolumn{2}{|c|}{Physical quantities in units of $L$} \\
 \hline
Hubble constant & $\mathcal{H}_0^{\text{phys}}  = \mathcal{H}_0^{\text{comp}}L^{-1}$\\
Conformal time & $\eta_{\rm phys}=\eta_{\rm comp} L $\\
Spatial coordinates & $x_{\rm phys}^i=x_{\rm comp}^i L$\\
Gravitational potential & $\phi_0^{\rm phys}= \phi_0^{\rm comp} L^0$\\
Velocity field & $\nabla \phi_0^{\rm phys}= \nabla \phi_0^{\rm comp} L^{-1}$\\
Density field & $\nabla^2 \phi_0^{\rm phys}= \nabla^2 \phi_0^{\rm comp} L^{-2}$\\
Frequency & $\omega^{phys}=\omega^{comp} L^{-1}$\\
Angular diameter distance &  $D_{\rm ang}^{phys}=D_{\rm ang}^{comp} L$\\
Parallax distance &  $D_{\rm par}^{phys}=D_{\rm par}^{comp} L$\\
Redshift drift & $\zeta^{phys}=\zeta^{comp} L^{-1}$\\
Redshift $z$ & dimensionless\\
Scale factor $a(\eta)$ & dimensionless\\
Growing mode $\mathcal{D}(\eta)$ & dimensionless\\
Cosmological parameters $\Omega_i$ & dimensionless\\
\hline
\end{tabular}
\end{table}

\section*{References}
\bibliographystyle{iopart-num}
\bibliography{ms.bib}

\end{document}